\documentstyle[graphicx,11pt]{report}
\newcommand{\be}{\begin{equation}}
\newcommand{\ee}{\end{equation}}
\newcommand{\bea}{\begin{eqnarray}}
\newcommand{\eea}{\end{eqnarray}}
\newcommand{\bd}{\begin{displaymath}}
\newcommand{\ed}{\end{displaymath}}
\newcommand{\bi}{\begin{itemize}}
\newcommand{\ei}{\end{itemize}}
\newcommand{\bc}{\begin{center}}
\newcommand{\ec}{\end{center}}
\newcommand{\bfl}{\begin{flushleft}}
\newcommand{\efl}{\end{flushleft}}
\newcommand{\bfr}{\begin{flushright}}
\newcommand{\efr}{\end{flushright}}

\def\6{\partial}

\def\={\!\!\!&=&\!\!\!}
\def\+{\!\!\!&&\!\!\!+~}
\def\-{\!\!\!&&\!\!\!-~}

\begin{document}

\pagenumbering{roman}

\begin{center}
{\large {''BABES-BOLYAI'' UNIVERSITY}, CLUJ, ROMANIA}

{\large Department of Theoretical Physics}

{\large \vspace{5cm} }

{\large \textbf{{\huge {Non-Fermi Behavior of the Strongly Correlated
Electron Systems}} }}

{\large \textbf{\vspace{2cm} }}

{\large \textbf{{\Large {Catalin Pascu Moca}} }}

{\large \textbf{\vspace{1cm} }}

{\large \textbf{{Adviser: Professor M.Crisan} }}

{\large \textbf{\vspace{3cm} }}

{\large \textbf{{Cluj - 1999} }}
\end{center}

\newpage\ 

\begin{center}
\textbf{{\Large {Abstract}} }

\textbf{{\large \textbf{\vspace{1cm}}} }
\end{center}

(1) The temperature dependence of the specific heat for a marginal Fermi
liquid has been calculated . We showed that the expected $T\ln T$ correction
is characteristic for the low temperature domain. The high temperature
domain has a supplementary correction. The results are in agreement with the
non-Fermi behavior of some metallic systems in the low temperature domain.

(2) We calculated the self-energy at $T=0$ for a two dimensional fermionic
system with hyperbolic dispersion. The existence of the saddle points in the
energy gives rise to a marginal behavior, a result which has been obtained
by numerical calculations.

(3) We present a simple demonstration that the two-dimensional fermionic
system with the energy $\varepsilon _{\mathbf{k}}=k_xk_y$ has a non-Fermi
behavior. The calculation of the wave function renormalization constant $Z$
were performed using the ''poor man's renormalization'' method and we have
showed that $Z\rightarrow 0$ with the infrared cut-off $\Lambda $ as $%
Z(\Lambda )=\Lambda ^\zeta $ where $\zeta $ is a constant.

(4) The electronic self-energy due to the electron-spin interaction is
calculated using the one-loop approximation for the two dimensional system
and quasi-two dimensional (anisotropic) model. We analyzed the relevance of
the diffusive modes and the temperature dependence of the magnetic
correlation length for a possible temperature dependence of the pseudogap.

(5) We study the influence of the amplitude fluctuations of a non-Fermi
superconductor on the energy spectrum of the two-dimensional Anderson
non-Fermi system. The classical fluctuations give a temperature dependence
in the pseudogap induced in the fermionic excitations.

(6) Using the field-theoretical methods we studied the evolution from BCS
description of a non-Fermi superconductor to that of Bose-Einstein
condensation (BEC) in one loop approximation. We showed that the repulsive
interaction between composite bosons is determined by the exponent $\alpha $
of the Anderson propagator in a two dimensional model. For $\alpha \neq 0$
the crossover is also continuous and for $\alpha =0$ we obtain the case of
the Fermi liquid.

(7) Using the renormalization group approach proposed by Millis for the
itinerant electron systems we calculated the specific heat coefficient $%
\gamma (T)$ for the magnetic fluctuations with susceptibility $\chi
^{-1}\sim \delta ^{\alpha }+\left| \omega \right| ^{\alpha }+f(q)$ near a
Lifshitz point. The constant value for $\alpha =4/5$ and the logarithmic
temperature dependence, specific heat for the non-Fermi behavior, have been
obtained in agreement with the experimental data.

\vspace{0.5cm}

This dissertation is based on the following papers:

\vspace{0.5cm}

$1.$ Specific Heat of a Marginal Fermi Liquid \textit{M.Crisan, C.P. Moca
Journal of Superconductivity }\textbf{9} 49 (1996)

$2.$ Marginal Behavior of a Two-Dimensional Fermionic System with Saddle
Points in the Energy \textit{C.P. Moca, M. Crisan Journal of
superconductivity} \textbf{10} 3 (1997)

$3.$ Non-Fermi Liquid Behavior of a System with Saddle Points in the Energy 
\textit{M. Crisan, C. P. Moca Modern Physics Letters} \textbf{B9} 1753 (1995)

$4.$ An Analytic Approach for the Pseudogap in the Spin Fluctuations Model 
\textit{\ C.P. Moca, I.Tifrea, M. Crisan (accepted for publication)}

$5.$ Electron-Fluctuation Interaction in a non-Fermi Superconductor \textit{%
M. Crisan, C.P. Moca, I. Tifrea Phys.Rev.} \textbf{B59} 14680 (1999)

$6.$ Field-theoretical Description of the Crossover between BCS and BEC in
an non-Fermi Superconductor \textit{C.P. Moca, I. Tifrea M. Crisan (accepted
for publication)}

$7.$ Renormalization Group Approach of the Itinerant Electron System Near a
Lifshitz Point \textit{C.P. Moca, I. Tifrea, M. Crisan (accepted for
publication)}

\newpage\ 

\textbf{{\Large {Acknowledgments}} }

\vspace{2.0cm}

Research presented in this thesis has been deeply influenced by my adviser
at ''Babes-Bolyai'' University, Prof. Mircea Crisan.

I owe a lot to my parents for their support and encouragement. Aside from
this dissertation there is a number of people whom I wish to thank. To my
colleagues Ionel Tifrea and Ioan Grosu for their support.

%\textbf{\newpage \ }\ 
\tableofcontents

\pagenumbering{arabic} %\newpage \ 

%\section{\textbf
\chapter{Non-Fermi Behavior and the Fermi Surface}

\vspace{0.1in}

One of the most important problems which was generated by the discovery of
cuprates superconductors is the nature of the normal phase. In fact the main
question is if a conventional Landau theory of the Fermi liquid (FL) can
describe this phase, or as was suggested by the experimental date, this
phase is in fact a special metal describe by a non-Fermi liquid (NFL).

The first model of a NFL was presented by Varma et.al.[1] and is well known
as the marginal-Fermi liquid (MFL) and developed by Kotliar et.al.[2] and
Littlewood et.al.[3]. This model is phenomenological but due to its
importance for the explanation of the experimental data much effort has been
done to give a microscopic foundation for the occurrence of the MFL
behavior. The main idea of the MFL model is the occurrence of a linear
energy dependence in the imaginary part of the self-energy due to the
anomalous frequency and temperature dependence of the density-density (
spin-spin ) response function. This particular form of the self-energy gives
rise to the pair-breaking effects, studied by Horbach et.al.[4] and Bendle
et.al.[5].

This particular behavior of the quasiparticle spectrum is essential for the
thermodynamics of the system, and the specific heat is one of the most
important physical quantities that has to be affected as mentioned by
Reizer[6] for the electrons interacting with a transverse electromagnetic
field.

For a three dimensional (3D) Fermi liquid, the imaginary part of the
self-energy has the form $Im\Sigma \left( \omega \right) \sim \omega ^2$ and
gives a logarithmic correction to the specific heat $C_V=\gamma T+\gamma
_{3D}T^3\ln T$. This behavior has been obtained in the framework of the
Landau theory, but this correction cannot describe the low-temperature
behavior of the $^3He$ which is considered as the standard Fermi liquid.

Anderson [7] showed that the low temperature dependence of the specific heat
for $^3He$ can be fitted by a $T\ln T$ dependence and suggested that this
dependence is given by coupling of the fermionic qausiparticles to the
collectives modes. Balian and Fredkin[8] and Berk and Schriffer[9] showed
that the coupling between the fermionic quasiparticles and the triplet
paramagnon gives a $T^3\ln T$ correction in the specific heat.

The correction to the linear dependence of the specific heat for a Fermi
liquid has been calculated by Amit et.al.[10] in terms of Landau parameters.

Carneiro and Petrick[11] performed a microscopic calculation of the
thermodynamic potential in terms of fully renormalized single-particle
propagator considering the effect of the quasiparticle lifetime. This method
can be successfully applied for the calculation of the specific heat of the
MFL [12]

\section{Thermodynamic Properties of the MFL}

\vspace{0.1in}

\subsection{Thermodynamic Potential and Entropy}

\vspace{0.1in}

In this section we calculate the specific heat of a MFL using a similar
method with that of Carneiro and Petrick [11]. The thermodynamic potential $%
\Omega $ is given by the fully renormalized single-particle propagator $%
G\left( \mathbf{p},\omega \right) $ as:

\begin{equation}
\Omega =T\sum_ne^{i\omega _n\eta }\left( -\ln \left[ -G^{-1}\left( \mathbf{p}%
,\omega _n\right) \right] -\Sigma \left( \mathbf{p,}\omega _n\right) G\left( 
\mathbf{p},\omega _n\right) \right)
\end{equation}
where the propagator $G\left( \mathbf{p},\omega _n\right) $ is given by
Dyson equation:

\begin{equation}
G^{-1}\left( \mathbf{p},\omega _n\right) =G_0^{-1}\left( \mathbf{p},\omega
_n\right) -\Sigma \left( \mathbf{p},\omega _n\right)
\end{equation}
$\Sigma \left( \mathbf{p},\omega _n\right) $ being the self-energy.

The thermodynamic potential $\Omega $ has the property that it is stationary
under variation of $G$ at fixed $G_0$ :

\begin{equation}
\left. \left( \frac{\delta \Omega }{\delta G}\right) \right| _{G_0}=0
\end{equation}
and shows that we can neglect the temperature dependence of the spectral
density $A\left( \mathbf{p},\omega \right) $.

\begin{equation}
A\left( \mathbf{p},\omega \right) =-2ImG\left( \mathbf{p},\omega \right)
\end{equation}
in the expression for the entropy $S\left( T\right) $ given by:

\begin{equation}
S\left( T\right) =-\left. \left( \frac{\delta \Omega }{\delta T}\right)
\right| _\mu
\end{equation}
Using:

\begin{equation}
G\left( \mathbf{p},\omega \right) =ReG\left( \mathbf{p},\omega \right)
-\frac i2A\left( \mathbf{p},\omega \right)
\end{equation}
and:

\begin{equation}
\Sigma \left( \mathbf{p},\omega \right) =Re\Sigma \left( \mathbf{p},\omega
\right) -\frac i2\Gamma \left( \mathbf{p},\omega \right)
\end{equation}
expression (1.5) becomes :

\[
S(T)=\sum_{\mathbf{p}}\int_{-\infty }^{+\infty }\frac{d\omega }{2\pi }\frac{%
\partial n\left( \omega \right) }{\partial T}2Im\ln \left[ -G^{-1}\left( 
\mathbf{p},\omega +i0^{+}\right) \right] 
\]
\begin{equation}
-\sum_{\mathbf{p}}\int_{-\infty }^{+\infty }\frac{d\omega }{2\pi }\frac{%
\partial n\left( \omega \right) }{\partial T}Re\Sigma \left( \mathbf{p}%
,\omega \right) A\left( \mathbf{p},\omega \right)
\end{equation}
\[
-\sum_{\mathbf{p}}\int_{-\infty }^{+\infty }\frac{d\omega }{2\pi }\frac{%
\partial n\left( \omega \right) }{\partial T}ReG\left( \mathbf{p},\omega
\right) \Gamma \left( \mathbf{p},\omega \right) 
\]
where $n\left( \omega \right) $ is the Fermi-Dirac distribution. This
equation has been written as:

\begin{equation}
S\left( T\right) =S_{DQ}\left( T\right) +S^{^{\prime }}\left( T\right)
\end{equation}
where $S_{DQ}\left( T\right) $ is the dynamical contribution expressed by:

\begin{equation}
S\left( T\right) =\sum_{\mathbf{p}}\int_{-\infty }^{+\infty }\frac{d\omega }{%
2\pi }\frac{\partial n\left( \omega \right) }{\partial T}\left( 2Im\ln
\left[ -G^{-1}\left( \mathbf{p},\omega +i0^{+}\right) \right] -ReG\left( 
\mathbf{p},\omega \right) \Gamma \left( \mathbf{p},\omega _n\right) \right)
\end{equation}
The second contribution has been identified as the contribution of terms
that has vanishing energy denominators.Equation (1.8) can be written as:

\begin{equation}
S\left( T\right) =S_0\left( T\right) +S_1\left( T\right) +S_2\left( T\right)
\end{equation}
where:

\begin{equation}
S_0(T)=-\sum_{\mathbf{p}}\left[ n(\varepsilon _{\mathbf{p}}^0)\ln
(n(\varepsilon _{\mathbf{p}}^0))+(1-n(\varepsilon _{\mathbf{p}}^0))\ln
(1-n(\varepsilon _{\mathbf{p}}^0))\right]
\end{equation}
and:

\begin{equation}
S_1(T)=-\sum_{\mathbf{p}}\int_{-\infty }^{+\infty }\frac{d\omega }{2\pi }%
\frac{\partial n\left( \omega \right) }{\partial T}Re\Sigma \left( \mathbf{p}%
,\omega \right) A\left( \mathbf{p},\omega \right)
\end{equation}
For $S_2(T)$ we get:

\begin{equation}
S_2(T)=-\sum_{\mathbf{p}}\int_{-\infty }^{+\infty }\frac{d\omega }{2\pi }%
\frac{\partial n\left( \omega \right) }{\partial T}\left[ \arctan B\left( 
\mathbf{p},\omega \right) +\frac{B\left( \mathbf{p},\omega \right) }{%
1+B^2\left( \mathbf{p},\omega \right) }\right]
\end{equation}
where:

\begin{equation}
B(\mathbf{p},\omega )=\frac{\Gamma \left( \mathbf{p},\omega \right) }{%
2ReG^{-1}\left( \mathbf{p},\omega \right) }
\end{equation}
and $\varepsilon _{\mathbf{p}}^0$ is the solution of the equation :

\begin{equation}
ReG^{-1}\left( \mathbf{p},\varepsilon _{\mathbf{p}}\right) =0
\end{equation}
If the damping of the quasiparticles is small, $S_0(T)$ is the entropy of
the Fermi liquid and gives a linear contribution in the specific heat of the
form:

\begin{equation}
C_V^0\left( T\right) =\gamma T
\end{equation}
In fact such a result will be obtained if we consider that in eq.(1.10) $%
G\left( \mathbf{p},\omega \right) =G_0\left( \mathbf{p},\omega \right) $
where $G_0(\mathbf{p},\omega )$ is the propagator for a FL, and in this
approximation $S_1(T)$ and $S_2(T)$ appears as corrections to the linear
dependence. These results will be used for the calculation of the specific
heat in the following section.

\vspace{0.1in}

\subsection{Specific Heat}

\vspace{0.1in}

A marginal Fermi liquid can be characterized by the functional form of the
frequency and temperature dependence of the single particle self-energy [1]:

\begin{equation}
\sum (\omega )=\left\{ 
\begin{array}{c}
\lambda \omega \ln \frac T{\omega _c}-i\frac \pi 2\lambda T\hspace{0.5in}%
\omega <T\hspace{0.2in} \\ 
\lambda \omega \ln \frac \omega {\omega _c}-i\frac \pi 2\lambda \omega %
\hspace{0.5in}T<\omega <\omega _c
\end{array}
\right.
\end{equation}
where $\lambda $ is a coupling constant and $\omega _c$ is an ultraviolet
cut-off.According to Varma and co-workers [1-3] the normal phase of high-$%
T_c $ oxides is an MFL that can be understood on the basis of an MFL
hypothesis i.e. the total electronic polarizability has a contribution $%
\mathcal{P}$ representing an electronic excitation with the property:

\begin{equation}
Im\mathcal{P}\left( \mathbf{p},\omega \right) =\left\{ 
\begin{array}{c}
-N(0)\frac \omega T\hspace{0.5in}\left| \omega \right| <T \\ 
-N(0)sgn(\omega )\hspace{0.5in}\left| \omega \right| >T
\end{array}
\right.
\end{equation}
for the leading frequency contribution, where $N\left( 0\right) $ is the
bare electronic density of states. All the universal anomalies as well as
the appearance of a 'Fermi surface' within the bound states can be
understood from this single hypothesis.

From eq. (1.18) a contribution to the resistivity proportional with $T$ is
obtained [1]. The nuclear relaxation time $T_1^{-1}(T)$, tunneling
conductance $g(V)$, optical conductivity $\sigma (\omega )$, fit well the
experimental data.

The electronic contribution to $C_V$, in the normal state cannot be
extracted from the data with high accuracy because the contribution of
phonons is very large above the high $T_C$ of these materials. $\kappa (T)$
in the normal state is observed to be nearly temperature independent[13]. If
the self-energy from the eq.(1.18) is used to calculate the entropy,
logarithmic correction to the linear-dependence are found. The observed $%
\kappa /T\sigma $ ratio allowed a weak correction.

The spectral density $A\left( \mathbf{p},\omega \right) $ can be calculated
from the eq.(1.4) and (1.18) as:

\begin{equation}
A\left( \mathbf{p},\omega \right) =\frac{\pi \lambda \omega }{\left[ \omega
-\varepsilon _{\mathbf{p}}-Re\Sigma (\omega )\right] ^2+\frac 14\pi
^2\lambda ^2\omega ^2}
\end{equation}
The contribution given by $S_1(T)$ has the form:

\begin{equation}
S_1(T)=\left\{ 
\begin{array}{c}
2N(0)T\left( F\left( \frac 12\right) -F\left( 0\right) \right) \ln \frac{%
\omega _c}T\hspace{0.5in}\omega <T\hspace{0.5in} \\ 
2N(0)T\left( F\left( \frac{\omega _c}{2T}\right) -F\left( \frac 12\right)
\right) \ln \frac{\omega _c}T\hspace{0.5in}\omega <T<\omega _c
\end{array}
\right.
\end{equation}

where $N(0)$ is the density of states and $F(x)$ is defined as:

\begin{equation}
F(x)=4x-x^2+x^2\tanh x+2\arctan \left( e^x\right) +2ReL_2(e^x)
\end{equation}
and:

\begin{equation}
L_2(x)=\int_0^x\frac{dt}t\ln \left| 1-t\right|
\end{equation}
The contribution given by $S_2(T)$ was calculated using eqs. (1.14), (1.18)
as:

\begin{equation}
S_2(T)=\left\{ 
\begin{array}{c}
2N(0)T\left( F\left( \frac 12\right) -F\left( 0\right) \right) \left(
1+\lambda \ln \frac{\omega _c}T\right) \hspace{0.5in}\omega <T\hspace{0.5in}
\\ 
2N(0)T\left( F\left( \frac{\omega _c}{2T}\right) -F\left( \frac 12\right)
\right) \left( 1+\lambda \ln \frac{\omega _c}T\right) \hspace{0.5in}\omega
<T<\omega _c
\end{array}
\right.
\end{equation}

Using the results for the entropy we can calculate the specific heat as:

\begin{equation}
C_V(T)=T\frac{\partial S(T)}{\partial T}
\end{equation}
In the following we will consider only the contribution given by the MFL, $%
S_0(T)$ giving a linear contribution specific to the FL.

>From eqs. (1.21) and (1.24) we get for $0<\omega <T:$

\begin{equation}
C_V=2N(0)T\left( F\left( \frac 12\right) -F(0)\right) \left( 1+2\lambda
\left( \ln \frac{\omega _c}T-1\right) \right)
\end{equation}
For the limit $T<\omega <\omega _c$ we get:

\begin{eqnarray}
C_V &=&2N(0)T\left( F\left( \frac{\omega _c}{2T}\right) -F\left( \frac
12\right) \right) \left( 1+2\lambda \left( \ln \frac{\omega _c}T-1\right)
\right) - \\
&&2N(0)T\left( \frac{\omega _c}{2T}\right) F^{\prime }\left( \frac{\omega _c%
}{2T}\right) \left( F\left( \frac{\omega _c}{2T}\right) -F\left( \frac
12\right) \right) \left( 1+2\lambda \left( \ln \frac{\omega _c}T-1\right)
\right)  \nonumber
\end{eqnarray}

where $F^{\prime }(x)=x^2/\cosh ^2(x).$

From eq.(1.26) we can see that for the MFL the temperature dependence has a
correction of the form $T\ln T$ but in the dominant region $T<\omega <\omega
_c$ a more complicated temperature dependence is added to this correction.

\vspace{0.1in}

\section{Marginal Behavior of a System with Saddle Points in the Energy}

\vspace{0.1in}

The non-Fermi behavior of electrons with hyperbolic dispersion $\varepsilon
_{\mathbf{k}}=k_xk_y$ has been predicted by News et.al. [14] in connection
with an electronic mechanism of superconductivity in the high critical
temperature superconductors. For such a two-dimensional (2D) system the
density of states shows a logarithmic energy dependence, and for a Fermi
energy $E_F$ near the saddle point ($E_c$) the critical temperature for the
superconducting state has been calculated, and is very sensitive to the new
energy scale $E^{*}$ given by the density-density excitation. The results
from [14] are not realistic because only the electron-hole channel has been
considered [15], but the results remain important for the explanation of the
anomalous properties of cuprate superconductors in the normal state. In
order to explain the physical properties of the normal state of these
superconductors we have to mention that the difficulties are due to the
study of a 2D interacting Fermi system and one of the simplest model is to
treat it as a Luttinger liquid, which exist for 1D Fermi liquid. However for
the (quasi)2D systems which are more realistic for cuprates superconductors,
there is no general demonstration for a MFL behavior. As it will be showed
the form of the Fermi surface is essential for the marginal behavior. A more
realistic model seems to be the model in which the Landau concept for the
quasiparticles is still valid and the deviations are due to the anomalous
scattering of electrons on the overdamped low energy excitations. Millis,
Monien and Pines [15] developed such a phenomenological form of the dynamic
susceptibility $\chi (\mathbf{q},\omega )$ which describe the normal and the
superconducting phase, the occurrence of the magnetic pseudogap and the
influence of the impurities. The pairing has also considered in this model,
but the Migdal Theorem discussed by Grosu and Crisan [16] is not valid in
the ''hot spots'' of the Fermi surface how was recently pointed by Amin and
Stamp [17]. Another interesting model has been proposed by Bernard et.al.
[18] and is based on the interaction between the electrons with a critical
bosonic mode. This interaction lead to a non-Fermi behavior (even to a Varma
et.al. [1] marginal behavior) and Crisan and Tataru [19] showed that this is
possible even in 3D case. The importance of the van-Hove singularities in
the density of states in the 2D Fermi systems for the cuprate
superconductors has been pointed by Friedel [20], Labbe and Bok [21] and in
a systematic study by the IBM group [22]. In one of their papers [23] it was
suggested that the saddle points singularities (called ''hot spots'') of the
Fermi surface are responsible for the deviations from the usual metallic
state. In the following we calculate the self energy of 2D model with
dispersion $\varepsilon _{\mathbf{k}}=k_xk_y/2m$ at $T=0$. Such a
calculation was also performed in [24] but the integrals were numerically
evaluated. We will give an analytic calculation which show the ''marginal
behavior'' of 2D systems.

\vspace{0.1in}

\subsection{Self-Energy of the 2D Fermi Systems}

\vspace{0.1in}

We consider the self-energy of a 2D electronic system interacting with the
electron-hole excitations which has the polarizability given in [23-24]:

\begin{equation}
Im\chi \left( \mathbf{q},z\right) =\frac 2{\pi \varepsilon _k}\left[ \left|
z+\varepsilon _{\mathbf{k}}\right| -\left| z-\varepsilon _{\mathbf{k}%
}\right| \right]
\end{equation}

\begin{equation}
Re\chi \left( \mathbf{q},z\right) =\frac 1{2\pi }\left[ \ln \left| \frac{%
4E_c\varepsilon _{\mathbf{k}}}{z^2-\varepsilon _{\mathbf{k}}^2}\right|
-\frac z{\varepsilon _{\mathbf{k}}}\ln \left| \frac{z+\varepsilon _{\mathbf{k%
}}}{z-\varepsilon _{\mathbf{k}}}\right| +2\right]
\end{equation}
where $\varepsilon _{\mathbf{k}}=k_xk_y/2m$ . The self-energy at $T=0$ is
given by the general equations:

\begin{equation}
Im\Sigma \left( \mathbf{p},\omega \right) =\frac{2u^2}\pi \int \frac{d^2%
\mathbf{k}}{(2\pi )^2}\int_0^\infty dzIm\chi \left( \mathbf{p-k},z\right)
ImG\left( \mathbf{k},z+\omega \right)
\end{equation}

\begin{equation}
Re\Sigma \left( \mathbf{p},\omega \right) =-\frac{2u^2}\pi \int \frac{d^2%
\mathbf{k}}{(2\pi )^2}\int_0^\infty dzRe\chi \left( \mathbf{p-k},z\right)
ImG\left( \mathbf{k},z+\omega \right)
\end{equation}
where $u$ is the electron-electron interaction which generate the bosonic
excitation. We mention that the dispersion

\begin{equation}
\varepsilon _{\mathbf{k}}=\frac 1{2m}(k_x^2-k_y^2)
\end{equation}
called the hyperbolic dispersion in [24] is equivalent to the simple case $%
\varepsilon _{\mathbf{k}}=k_xk_y/2m$. In order to perform the calculations
using eqs.(1.28) and (1.29) we will perform the transformation:

\begin{equation}
k_x=\frac k2\left( \lambda +\frac 1\lambda \right)
\end{equation}

\begin{equation}
k_y=\frac k2\left( \lambda -\frac 1\lambda \right)
\end{equation}
and the dispersion has the form

\begin{equation}
\varepsilon _{\mathbf{k}}=\frac{k_xk_y}{2m}
\end{equation}

\vspace{0.1in}

\subsection{Imaginary Part of the Self-Energy}

\vspace{0.1in}

Equation (1.30) can be transformed using:

\begin{equation}
ImG(\mathbf{k},\omega +z)=-\pi \delta (z+\omega -\varepsilon _{\mathbf{k}})
\end{equation}
as:

\begin{equation}
Im\Sigma \left( \mathbf{p},\omega \right) =\frac{u^2}{\pi ^3}\int_0^\infty
kdk\int_{-\infty }^\infty \frac{d\lambda }\lambda \frac 1{\varepsilon _{%
\mathbf{p-k}}}\left[ \left| \varepsilon _{\mathbf{k}}-\omega +\varepsilon _{%
\mathbf{p-k}}\right| -\left| \varepsilon _{\mathbf{k}}-\omega -\varepsilon _{%
\mathbf{p-k}}\right| \right]
\end{equation}

Let us calculate this integral first in the limit $\omega \ll \varepsilon _k$%
. In this case eq.(1.37) becomes:

\begin{equation}
Im\Sigma \left( \mathbf{p},\omega \right) =\frac{u^2}{\pi ^3}\int_0^\infty
kdk\left[ \varepsilon _{\mathbf{k}}-\omega \right] \int_{-\infty }^\infty 
\frac{d\lambda }\lambda \frac{2m}{(p_x^2+k_x^2-2p_xk_x)-(p_y^2+k_y^2-2p_yk_y)%
}
\end{equation}
and if we take the $\mathbf{p}=(p,0)$ this equation becomes:

\begin{equation}
Im\Sigma \left( \mathbf{p},\omega \right) =\frac{2mu^2}{\pi ^3p}%
\int_0^\infty kdk\left[ \varepsilon _{\mathbf{k}}-\omega \right]
\int_{-\infty }^\infty \frac{d\lambda }{\lambda ^2-(p^2-k^2)/(pk)+1}
\end{equation}
and performing the integration over $\lambda $ we get:

\begin{eqnarray}
Im\Sigma \left( \mathbf{p},\omega \right) &=&\frac{2mu^2}{\pi ^3p}%
\int_0^\infty kdk\left[ \frac{k^2}{2m}-\omega \right] \times \\
&&\left[ \frac{2pk}{p^2-k^2}arcth\frac{2pk-p^2-k^2}{p^2-k^2}+\frac{2pk}{%
p^2-k^2}arcth\frac{p^2+k^2}{p^2-k^2}\right]  \nonumber
\end{eqnarray}
In the approximation $arcth(x)\cong x$ we calculate the first contribution
in eq.(1.40) as:

\begin{equation}
Im\Sigma _1(\mathbf{p},\omega )=-\frac{2mu^2}{\pi ^3p}\int_0^\infty dk\left[ 
\frac{k^2}{2m}-\omega \right] \frac{2pk}{(p+k)^2}
\end{equation}
and for $k\ll p$ eq.(1.41) will be written as:

\begin{equation}
Im\Sigma _1(\mathbf{p},\omega )=-\frac{2mu^2}{\pi ^3p^2}\int_0^\infty
kdk\left[ \frac{k^2}{2m}-\omega \right]
\end{equation}
where we neglected the contribution given by $k^2+2pk\ll p^2$ in the
denominator of eq.(1.41). From eq.(1.42) we obtain:

\begin{equation}
Im\Sigma _1(\mathbf{p},\omega )=\frac{mu^2}{\pi ^3}\left[ \frac{p^2}{2m}%
-\omega \right]
\end{equation}

In the same approximation we calculate the contribution from the second term
of eq. (1.40) and:

\begin{equation}
Im\Sigma _2(\mathbf{p},\omega )=\frac{2mu^2}{\pi ^3}\left[ \frac{p^2}{2m}%
-\omega \right]
\end{equation}

From these equations the imaginary part of the self-energy becomes:

\begin{equation}
Im\Sigma (\mathbf{p},\omega )=-\frac{3mu^2}{\pi ^3}\left[ \omega -\frac{p^2}{%
2m}\right]
\end{equation}
and for $p=p_F$ we get:

\begin{equation}
Im\Sigma (\mathbf{p},\omega )=-3N(0)\left( \frac U\pi \right) ^2\left(
\omega -E_F\right)
\end{equation}
where $N(0)=m/\pi $ and $E_F=p_F^2/2m$

\vspace{0.1in}

\subsection{Real Part of the Self-Energy}

\vspace{0.1in}

The general expression for $Re\Sigma (\mathbf{p},\omega )$ in the second
order of perturbation theory has the general form:

\begin{equation}
Re\Sigma (\mathbf{p},\omega )=-\frac{u^2}\pi \int \frac{d^2\mathbf{k}}{(2\pi
)^2}\int_0^\infty dzRe\chi (\mathbf{p-k},z)ImG(\mathbf{k},z+\omega )
\end{equation}
where $ImG$ will be considered as $\delta $-function and $Re\chi (\mathbf{k}%
,z)$ is given by eq.(1.29). If we perform the integral over $z$ eq.(1.47)
becomes:

\[
Re\Sigma (\mathbf{p},\omega )=-\frac{u^2}{4\pi ^4}\int_0^\infty kdk\int 
\frac{d\lambda }\lambda \left[ \ln \left| \frac{4E_c\varepsilon _{\mathbf{p-k%
}}}{\left( \varepsilon _{\mathbf{k}}-\omega \right) ^2}-\varepsilon _{%
\mathbf{p-k}}^2\right| \right. 
\]

\begin{equation}
\left. -\frac{\varepsilon _{\mathbf{k}}-\omega }{\varepsilon _{\mathbf{p-k}}}%
\ln \left| \frac{\varepsilon _{\mathbf{k}}-\omega +\varepsilon _{\mathbf{p-k}%
}}{\varepsilon _{\mathbf{k}}-\omega -\varepsilon _{\mathbf{p-k}}}\right|
+2\right] \theta (\varpi -\varepsilon _{\mathbf{k}}+\omega )
\end{equation}
where $\varpi $ is a frequency cut-off. The integral containing the first
two terms will be transformed as:

\[
Re\Sigma (\mathbf{p},\omega )=-\frac{u^2}{4\pi ^4}\int_0^\infty kdk\int 
\frac{d\lambda }\lambda \left[ 1-\frac{\varepsilon _{\mathbf{k}}-\omega }{%
\varepsilon _{\mathbf{p-k}}}\ln \left| 1-\frac{\varepsilon _{\mathbf{k}%
}-\omega }{\varepsilon _{\mathbf{p-k}}}\right| \right. 
\]

\begin{equation}
\left. -1+\frac{\varepsilon _{\mathbf{k}}-\omega }{\varepsilon _{\mathbf{p-k}%
}}\ln \left| 1+\frac{\varepsilon _{\mathbf{k}}-\omega }{\varepsilon _{%
\mathbf{p-k}}}\right| -\ln \frac{\varepsilon _{\mathbf{p-k}}}{4E_c}\right]
\end{equation}

Using the expansion:

\begin{equation}
\ln (1+x)=x-\frac{x^2}2
\end{equation}
we get:

\begin{equation}
Re\Sigma _1(\mathbf{p},\omega )=-\frac{u^2}{4\pi ^4}\int_0^\infty kdk\int 
\frac{d\lambda }\lambda \left[ \frac{\varepsilon _{\mathbf{k}}-\omega }{%
\varepsilon _{\mathbf{p-k}}}+\ln \left| \frac{4E_c}{\varepsilon _{\mathbf{p-k%
}}}\right| \right]
\end{equation}
In order to perform the integration over $\lambda $ we calculate

\begin{equation}
\varepsilon _{\mathbf{p-k}}=p^2+k^2-2\mathbf{pk}
\end{equation}

\begin{equation}
\mathbf{pk}=\frac{pk}2\left( \lambda +\frac 1\lambda \right)
\end{equation}
where we take ($2m=1$) and (1.51) becomes:

\[
Re\Sigma _1(\mathbf{p},\omega )=-\frac{u^2}{4\pi ^3}\int_0^\infty kdk\frac{%
\varepsilon _{\mathbf{k}}-\omega }{\left| (p-k)(p+k)\right| }\theta (\varpi
+\omega -\varepsilon _{\mathbf{k}}) 
\]

\begin{equation}
=-\frac{u^2}{8\pi ^3}\int_0^{\omega +\varpi }dk^2\left[ \left( 1+\frac{%
p^2-\omega }{k^2-p^2}\right) +\frac 4{k^2-p^2}\right]
\end{equation}
which gives:

\begin{equation}
Re\Sigma _1(\mathbf{p},\omega )=\frac{u^2}{8\pi ^3}\left[ (\varpi +\omega
)+(\varepsilon _{\mathbf{p}}-\omega )\ln \left| \frac{\varpi +\omega }{%
\varepsilon _{\mathbf{p}}}-1\right| \right] -\frac{u^2}{2\pi ^3}\ln \left| 
\frac{\varpi +\omega }{\varepsilon _{\mathbf{p}}}-1\right|
\end{equation}

The last term of eq. (1.49) is:

\begin{equation}
Re\Sigma _2(\mathbf{p},\omega )=-\frac{u^2}{2\pi ^2}\int_0^\infty kdk\int 
\frac{d\lambda }\lambda \theta (\varpi -\varepsilon _{\mathbf{k}}+\omega )
\end{equation}
we have to consider $k^2<\varpi +\omega $ and using:

\begin{equation}
\lambda =\frac{2(k_x+k_y)}k\simeq \frac{2(k_x+k_y)}{\sqrt{\varpi +\omega }}
\end{equation}
we obtain that $Re\Sigma _2(\mathbf{p},\omega )$ vanished for $%
p-p_F\longrightarrow 0$. The final form of the real part of the self-energy
is:

\begin{equation}
Re\Sigma (\mathbf{p},\omega )=\frac{u^2}{8\pi ^3}\left[ (\varpi +\omega
)+(E_F-\omega )\ln \left| \frac \omega {E_F}-1\right| \right] -\frac{u^2}{%
2\pi ^3}\ln \left| \frac \omega {E_F}-1\right|
\end{equation}

We can see that the divergence in the real part has the form:

\begin{equation}
Re\Sigma (\mathbf{p},\omega )=(\omega -E_F)\ln \left| \frac \omega
{E_F}-1\right|
\end{equation}

These results have been obtained in the one loop approximation. In the
following we will show that this model present a non-Fermi behavior using
the renormalization group method.

\section{''Poor Man's Renormalization'' for a System with Saddle Points
in the Energy}

\vspace{0.1in}

\subsection{Model}

\vspace{0.1in}

Using the ''poor man's renormalization '' we will present a direct
demonstration that the saddle points can give the non-Fermi liquid behavior.

We consider, following Newns et.al. [14] a 2D fermionic system with the
energy:

\begin{equation}
\varepsilon _{\mathbf{k}}=k_xk_y
\end{equation}
where $\left| k_x\right| <k_c;\left| k_y\right| <k_c,$ $k_c$ being the
cut-off. If the Fermi level is closed to the van-Hove singularity the nearly
2D fermionic system present two important differences to the normal metal.
The first one is that the phase space for scattering is much less restricted
that in the conventional metal and the lifetime of the quasiparticles
becomes of the order of their energy. Such a fermionic system becomes a MFL.
The second point is that the presence of the van-Hove singularity close to
the Fermi energy is destroying the nesting of the Fermi surface which gives
magnetic instabilities which are competing the Cooper instability. The
effective interaction between the quasiparticles has been calculated in
ref.[14] as:

\begin{equation}
V_{eff}=\frac 1{\Pi (\mathbf{q},\omega )}
\end{equation}
where the polarizability $\Pi (\mathbf{q},\omega )$ is given by:

\begin{equation}
\Pi (\mathbf{q},z)=\frac f{2D}\left[ P\left( z,\frac{q_x}{q_c}\right)
+P\left( z,\frac{q_y}{q_c}\right) \right]
\end{equation}
where $f$ is a fraction of the area of the Brillouin zone and $D=k_c^2$. The
function $P(z,x)$ is defined as:

\begin{equation}
P(z,x)=g(1)-g(-1)-g\left( \frac 1x\right) +g\left( -\frac 1x\right) -g\left(
1-\frac 1x\right) +g\left( -1+\frac 1x\right)
\end{equation}
where

\begin{equation}
g(u)=(z+u)\ln \left| z+u\right|
\end{equation}

\begin{equation}
z=\frac{\left| \nu \right| }{\varepsilon _q}
\end{equation}

Using in eq. (1.63) the expansion:

\begin{equation}
\ln x\cong 2\frac{x-1}{x+1}
\end{equation}
we can approximate (1.62) by the expression:

\begin{equation}
P(z,x)=-\frac{4z}{x^2}+\frac 4x-\frac 2{x^2}
\end{equation}
the effective potential has the form

\begin{equation}
V_{eff}(\mathbf{q},\nu )=\left\{ 
\begin{tabular}{ll}
$\frac{Z_e}f\left[ V_0+A\left| \nu \right| \frac{\sqrt{D}}{\sqrt{\varepsilon
_{\mathbf{q}}}}\right] $ & \hspace{0.5in}$\nu <\sqrt{E^{*}\varepsilon _{%
\mathbf{q}}}$ \\ 
$V_c$ & \hspace{0.5in}$\nu >\sqrt{E^{*}\varepsilon _{\mathbf{q}}}$%
\end{tabular}
\right.
\end{equation}
where $V_0$ is given as $V_0=\frac{4D}{Z_e},Z_e$ and A are constants, and $%
V_c$ is the saturation value of $V_{eff}(\mathbf{q},\nu )$ in the high
frequency region and $E^{*}=\left[ fV_c-Z_eV_0\right] ^2/D$ .

\vspace{0.1in}

\subsection{Self-Energy and the Renormalization Constant $Z$}

\vspace{0.1in}

We start with the general expression for the self-energy:

\begin{equation}
\Sigma (\mathbf{k},\omega )=i\int \frac{d^2\mathbf{q}}{(2\pi )^2}\int \frac{%
d\nu }{2\pi }V_{eff}\left( \mathbf{q},\nu \right) G_0\left( \mathbf{k-q}%
,\omega -\nu \right)
\end{equation}
where $G_0$ is the Green function for the free fermionic particles and $%
V_{eff}$ is given by eq.(1.61) The integral over $\nu $ will be written as:

\[
I(\mathbf{k-q},\omega )=i\int_{-\infty }^\infty \frac{d\nu }{2\pi }\frac{V_c%
}{\omega -\nu -\varepsilon _{\mathbf{k-q}}+i\delta } 
\]

\begin{equation}
+i\int_{-\nu _0}^{\nu _0}\frac{d\nu }{2\pi }\frac{V_{1-}V_c}{\omega -\nu
-\varepsilon _{\mathbf{k-q}}+i\delta }
\end{equation}

\[
+i\int_{-\nu _0}^{\nu _0}\frac{d\nu }{2\pi }\frac 1{\sqrt{\varepsilon _{%
\mathbf{q}}}}\frac 1{\omega -\nu -\varepsilon _{\mathbf{k-q}}+i\delta } 
\]
where $\nu _0=\sqrt{E^{*}\varepsilon _{\mathbf{q}}}$ and $E^{*}$ is the
matching energy. In order to calculate $Z$ the renormalization will be
performed only on the real part of the self-energy $\Sigma ^R(\mathbf{k}%
,\omega )=Re\Sigma (\mathbf{k},\omega )$ and we will be interested only in
the contribution:

\[
ReI\left( \mathbf{k-q},\omega \right) =-V_c\theta \left( \nu _0-\left|
\omega -\varepsilon _{\mathbf{k-q}}\right| \right) 
\]

\begin{equation}
+\left( V_1-V_c\right) \theta \left( \nu _0-\left| \omega -\varepsilon _{%
\mathbf{k-q}}\right| \right)
\end{equation}

\[
+A\frac{\left| \omega -\varepsilon _{\mathbf{k-q}}\right| }{\sqrt{%
\varepsilon _{\mathbf{q}}}}\theta \left( \nu _0-\left| \omega -\varepsilon _{%
\mathbf{k-q}}\right| \right) 
\]
where $\theta (x)$ is the step function.

Using this result the real part of the self-energy (1.69) is:

\begin{equation}
\Sigma ^R(\mathbf{k},\omega )=\Sigma _1^R(\mathbf{k},\omega )+A\int \frac{d^2%
\mathbf{q}}{(2\pi )^2}\frac{\left| \omega -\varepsilon _{\mathbf{k-q}%
}\right| }{\sqrt{\varepsilon _{\mathbf{q}}}}\theta \left( \nu _0-\left|
\omega -\varepsilon _{\mathbf{k-q}}\right| \right)
\end{equation}
where $\Sigma _1^R(\mathbf{k},\omega )$ gives no contribution to the
renormalized part. In order to calculate the renormalization constant $Z$ we
will use the ''poor man's renormalization '' taking $\varepsilon _q$ as
shown in Fig.1.1.

\begin{figure}[tbh]
\centering
\includegraphics[clip,width=0.8\textwidth]{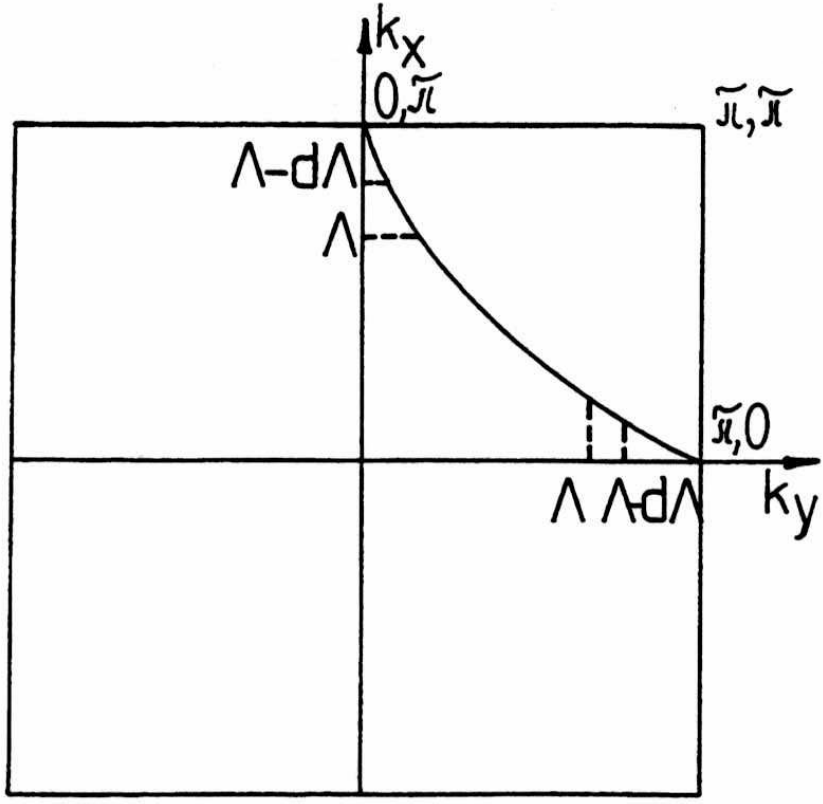}
\caption{The domain of integration for
the calculation of self-energy}
\label{Fig1}
\end{figure}
%\FRAME{ftbpFU}{3.2993in}{2.9196in}{0pt}{\Qcb{The domain of integration for
%the calculation of self-energy}}{}{fig1.jpg}{\special{language "Scientific
%Word";type "GRAPHIC";maintain-aspect-ratio TRUE;display "PICT";valid_file
%"F";width 3.2993in;height 2.9196in;depth 0pt;cropleft "-0.0030";croptop
%"1.0111";cropright "0.9983";cropbottom "0.0101";filename
%'D:/TEZA/FIG1.JPG';file-properties "XNPEU";}}

If we introduce near $k_c$ the cut-off $\Lambda $ from eq. (1.72) we
calculate $d\Sigma ^R$ as:

\[
d\Sigma ^R\left( \Lambda ,\mathbf{k},\omega \right) =\frac A2\int_{\Lambda
-d\Lambda }^\Lambda \frac{dq_x}{2\pi }\int_0^\Lambda \frac{dq_y}{2\pi }\frac{%
\left| \omega -\varepsilon _{\mathbf{k-q}}\right| }{\sqrt{\varepsilon _{%
\mathbf{q}}}}\delta \left( q_y-\frac{k_c^2}{q_x}\right) \frac 1{q_x}+ 
\]

\begin{equation}
+\frac A2\int_{\Lambda -d\Lambda }^\Lambda \frac{dq_y}{2\pi }\int_0^\Lambda 
\frac{dq_x}{2\pi }\frac{\left| \omega -\varepsilon _{\mathbf{k-q}}\right| }{%
\sqrt{\varepsilon _{\mathbf{q}}}}\delta \left( q_x-\frac{k_c^2}{q_y}\right)
\frac 1{q_y}
\end{equation}
which gives

\begin{equation}
d\Sigma ^R\left( \Lambda ,\mathbf{k},\omega \right) =-A\frac{d\Lambda }%
\Lambda \omega +Av_Fk\frac{d\Lambda }\Lambda
\end{equation}
where:

\begin{equation}
d\Sigma ^R(\Lambda ,\mathbf{k},\omega )=\Sigma ^R(\Lambda ,\mathbf{k},\omega
)-\Sigma ^R(\Lambda -d\Lambda ,\mathbf{k},\omega )
\end{equation}

In order to obtain the general equation of the renormalized quantities we
write the propagator $G(\Lambda ,\mathbf{k},\omega )$ as:

\begin{equation}
G(\Lambda ,\mathbf{k},\omega )=\frac{Z(\Lambda )}{\omega -v_F(\Lambda
)k+i\gamma }
\end{equation}
and:

\begin{equation}
G(\Lambda ,\mathbf{k},\omega )=\frac{Z(\Lambda )}{\omega -v_F(\Lambda
)k+\Sigma ^R(\Lambda ,\mathbf{k},\omega )+i\Sigma ^I(\Lambda ,\mathbf{k}%
,\omega )}
\end{equation}
where we approximated $\varepsilon (k)=v_Fk$ , $k$ being the momentum
measured from the Fermi surface momentum $k_F$. From these two equations we
get:

\begin{equation}
dZ(\Lambda )\left( \omega -v_F(\Lambda )k\right) -dZ\left( \Lambda \right)
\Sigma ^R(\Lambda ,\mathbf{k},\omega )-Z(\Lambda )\delta \Sigma ^R(\Lambda ,%
\mathbf{k},\omega )=-dv_F(\Lambda )k
\end{equation}
and in this equation the second term can be neglected since it given no
contribution to the renormalized quantities and it is in fact a constant
which can be absorbed in $\omega $. Therefore eq.(1.78) becomes:

\begin{equation}
\frac{dZ(\Lambda )}{Z(\Lambda )}\left( \omega -v_Fk\right) -d\Sigma
^R(\Lambda )=-dv_F(\Lambda )k
\end{equation}
on the other hand if $d\sum^R(\Lambda )$ has the general form:

\begin{equation}
-d\Sigma ^R(\Lambda )=C_1\frac{d\Lambda }\Lambda \omega +C_2v_Fk\frac{%
d\Lambda }\Lambda
\end{equation}
we get the general equations:

\begin{equation}
\frac 1{Z(\Lambda )}\frac{dZ(\Lambda )}{d\Lambda }=\frac{C_1}\Lambda
\end{equation}
and

\begin{equation}
\frac 1{v_F(\Lambda )}\frac{dv_F(\Lambda )}{d\Lambda }=\frac{C_1-C_2}\Lambda
\end{equation}

>From the eqs.(1.81)-(1.82) we get:

\begin{equation}
v_F(\Lambda )=0
\end{equation}
and

\begin{equation}
Z(\Lambda )=\Lambda ^\zeta
\end{equation}

Equation (1.83) shows that the Fermi velocity is constant and Eq.(1.84)
shows the important result :

\begin{equation}
\lim\limits_{\Lambda \rightarrow 0}Z(\Lambda )=0
\end{equation}
with an exponent

\begin{equation}
\zeta =\frac{Z_eA\sqrt{D}}{fk_c}
\end{equation}
and using the relation $D=k_c^2$ we obtain the exponent

\begin{equation}
\zeta =\frac{Z_e\Lambda }f
\end{equation}
which depend only on the constants contained in the effective potential.

\vspace{0.1in}

\section{Summary of the Results}

\vspace{0.1in}

$\bullet $In the first part of this chapter we performed an analytic
calculation of the electronic specific heat of MFL and we showed that the $%
T\ln \frac{\omega _c}T$ correction predicted in [3] appears only in the
domain $0<\omega <T$. For energies in the domain $T<\omega <\omega _c$ a
supplementary correction depending also on $\ln \frac{\omega _c}T$ appears.

Following the same method in the nonmarginal case the correction in $C_V$ is 
$T^3\ln T$ if the electron-electron or electron-phonon interactions are
considered for 3D FL model [26]. This behavior makes very difficult the
separation of MFL contribution as mentioned by Varma et.al.[1-2]. Recently
Coffey and Bedell [27] showed that a 2D Fermi liquid has analytic correction
in $T$ of the form $\delta C_V^{2D}=\gamma _{2D}T^2$ and there is no
breakdown of the Fermi liquid in 2D. On the other hand we mention that for
the uranium compound $U_{0.2}Y_{0.8}Pd_3$ [28] was found a $T\ln T$
correction in the specific heat, a result that shows the existence of the
non-Fermi behavior in the real systems.

$\bullet $In the second part of this chapter we showed analytically that the
2D electronic system with hyperbolic dispersion exhibits the marginal
behavior ( $Im\sum (p_F,\omega )\sim \omega $) obtained numerically in [24].
The marginal behavior is given by the saddle points of the Fermi surface as
was predicted in [23]. We have to mention that the marginal phenomenological
model is not restricted to the 2D Fermi systems but the cuprates
superconductors are (quasi)2D systems.

$\bullet $ Using the ''poor man's renormalization '' method we showed in the
last part of this chapter the occurrence of the MFL behavior for 2D
fermionic systems with saddle points in the energy. Finally we mention that
the NFL behavior is typically for the 1D fermionic systems (Luttinger liquid
). The 2D fermionic systems with saddle points in the energy is very
realistic for the cuprates superconductors, but cannot be treated
analytically without renormalization group methods. More than that even the
NFL models based on the coupling to a gauge field [29] or on the superlong
range interaction [30] has been treated by this method.

\chapter{Anderson non-Fermi Model}

\vspace{0.1in}

The microscopic description of the superconducting state in cuprates
superconductors is a very difficult problem because at the present time it
is generally accepted that in the normal state the elementary excitations
are not describe by the Fermi liquid theory. However using the BCS-like
pairing model the Gorgov equations have been applied to describe the
superconducting state in the hypothesis that the normal state is a non-Fermi
liquid described by the Anderson model [1]. The superconducting state
properties have been discussed by different authors [2-8] and even if these
descriptions are phenomenological, it can be a valid starting point for a
microscopic model. Recent experimental data (ARPES) showed that these
materials presents even more remarkable deviations from the Fermi-liquid
behavior due to the occurrence of the pseudogap at the Fermi-surface.

The occurrence of the pseudogap has been explained using different concepts
such as the spin fluctuation [9], preformed pairs [10], SO(5) symmetry [11],
spin-charge separation [12], and the fluctuation of the order parameter
induced pseudogap [13].

\vspace{0.1in}

\section{The Model}

\vspace{0.1in}

The non-Fermi liquid behavior of the normal state for the cuprates
superconductors proposed by Anderson [1] was developed by different authors
in order to describe the superconducting state in the framework of the BCS
theory. In the normal state the electrons are describe by the Green function:

\begin{equation}
G\left( \mathbf{k},i\omega _n\right) =\frac{g(\alpha )e^{i\phi }}{\omega
_c^\alpha \left( i\omega _n-\varepsilon _{\mathbf{k}}\right) ^{1-\alpha }}
\end{equation}
where $\omega _c$ is a cutoff energy and $0<\alpha <1$ and $g(\alpha )=\pi
\alpha /2\sin \left( \frac{\pi \alpha }2\right) $ and $\phi =-\pi \alpha /2$%
. The Green function given by eq.(2.1) was given by Abrahams [14] and used
by Yin and Chakravarty [15] to study the non-Fermi liquid superconductors.
However, the time reverse symmetry is violated unless $\phi =-\pi \alpha /2$%
. This is because of the invariance with respect to $\omega \rightarrow
-\omega $ and $\varepsilon _{\mathbf{k}}\rightarrow -\varepsilon _{\mathbf{k}%
}$. When $\varepsilon _{\mathbf{k}}=0$, the spectral function is similar to
that of Luttinger liquid without spin-charge separation in $d=1$ [16] but is
very different when $\varepsilon _{\mathbf{k}}\neq 0$. To preserve the
equal-time anticommutation relation of the fermions, the spectral function
must satisfy the sum rule $\int_{-\infty }^\infty d\omega A(\mathbf{k}%
,\omega )=1$, from which relation it can be found the value of $g(\alpha )$
given above. With the above value of $g(\alpha )$ the anticommutation
relations will be satisfied if $\left| \varepsilon _{\mathbf{k}}\right|
<\omega _c$, which is precisely the regime in which a scaling theory is
appropriate. The density of states can be calculated from the spectral
function [15]. In contrast to the Luttinger liquid the density of states
does not vanished as $\omega \rightarrow 0$. In fact it is unchanged from
the Fermi liquid value as $\alpha \rightarrow 0$. It would agree exactly
with the Fermi liquid value, if $\omega _c$ is chosen to be $W$(the band
width)

The free action for the model is:

\begin{equation}
S=\int d\omega d^2\mathbf{p}G^{-1}(\mathbf{p},\omega )\Psi _{\mathbf{p}%
,\sigma }^{+}(\omega )\Psi _{\mathbf{p},\sigma }(\omega )
\end{equation}
where the Green function $G$ is given by eq.(2.1). If we follow Shankar [17]
we see that the four fermions interaction is irrelevant. The spectral
anomaly is more stable than the Fermi liquid. In fact in the weak coupling
regime it does not even allow a superconducting instability [15]. The
coupling has to reach a treashold before superconducting instability occurs.

\vspace{0.1in}

\section{A Novel Feature of the non-Fermi Behavior. The Occurrence of the
Pseudogap.}

\vspace{0.1in}

The recent angle-resolved photoemission spectroscopy (ARPES) experiments
confirmed the occurrence of pseudogap in the density of states of the
electronic excitations from the normal and superconducting phase of the
cuprate superconductors.

Many theoretical approaches and Monte-Carlo simulations [18-19] of strongly
correlated electron systems has been performed in order to explain the
experimental data. Some common feature are observed in many of these models,
the most important being the fact that the origin of the pseudogap is the
interaction between the electronic excitations and fluctuations. This idea
leads to the hypothesis that the pseudogap appears in the proximity of a
quantum phase transition (QPT) [20-21]

A consistent picture is emerging from the study of the Hubbard model in $2D$
and $3D$ using Monte-Carlo calculations and the analytically
non-perturbative many-body approach perform by Tremblay group. The main
results obtain by authors in Ref.[22-25] is the occurrence of a temperature
independent pseudogap for a 2D Hubbard model taking the interaction between
the electrons and the renormalized classical fluctuations. In this regime
the coherence length is increasing with decreasing of the temperature and a
good approach for the model is a QPT.

On the other hand the spin fluctuations are present in both normal and
superconducting phase of the cuprates. The Fermi surface evolution and the
temperature dependence of the pseudogap have been studied by Chubukov
[26-27] using the phenomenological model proposed by Millis Monien and Pines
[28] (MMP).

In this section we will perform an analytic calculation of the pseudogap
given by the interaction between electrons and magnetic fluctuations.
Starting such a problem we expect that a tractable approach will be only
one-loop diagram for the self-energy. This approximation does not take the
vertex corrections which are important because in such a system probably the
Migdal theorem is not valid. However, even such a simple approximation can
gives us an idea if the interaction between electrons and spin fluctuations
are good candidate to explain pseudogap.

The self-energy is essential determined in the model by the dynamic
susceptibility $\chi (\mathbf{k},\omega )$ which will be discussed below.
Such a discussion is also usefully because at the present time there are
models with $\chi (\mathbf{k},\omega )$ pure diffusive [29] or $\chi (%
\mathbf{k},\omega )$ describing overdamped fluctuations. Recently it was
presented a model with both contributions in $\chi (\mathbf{k},\omega )$
[30].

\vspace{0.1in}

\subsection{Dynamic Susceptibility}

\vspace{0.1in}

We will consider that the electronic system interacts with magnetic
excitations near a quantum critical point (QCP) determined by the condition $%
J(\mathbf{Q})=-\chi _0^{-1}(\mathbf{Q})$, where $J(\mathbf{Q})$ is the
magnetic interaction and $\chi _0^{-1}(\mathbf{Q})$ is the bare
susceptibility in $\mathbf{Q}=(\pm \pi ,\pm \pi )$. The general form for $%
\chi (\mathbf{q},\mathbf{\omega })$ describing the overdamped
antiferromagnetic fluctuations is:

\begin{equation}
\chi (\mathbf{q},\omega )=\frac{\chi _0(\mathbf{Q})}{\delta +a^2q^2-i\gamma
\omega }
\end{equation}
where $a$ is of the order of the lattice spacing, $\delta $ is a parameter
characterizing the distance from the QCP and $\gamma =\frac 1{\omega _{SF}}$%
, $\omega _{SF}$ being the characteristic energy of the spin fluctuations.
The renormalization group method (RNG) has been applied to the analysis of
the phase transition of the itinerant electronic system by Millis [31]
taking into consideration the dynamic critical exponent $z$. The correlation
time of the order parameter $\tau $ and the correlation length $\xi $ are
connected by:

\begin{equation}
\tau =\xi ^z
\end{equation}
which shows that $\tau $ diverges more than $\xi $. From eqs.(2.3) and (2.4)
we get:

\begin{equation}
\xi ^2=\frac{a^2}\delta
\end{equation}

\begin{equation}
\tau =\frac 1{\gamma \delta }=\frac{\omega _{SF}}\delta
\end{equation}
so the $z=2$, one of the most important predictions obtained in the weak
coupling approach. At $T=0$ (i.e. for the quantum phase transition) the
static and dynamics are mixed in contrast with the case $T>0$ where the
dynamic behavior affects the static behavior.

The presence of the particle-hole continuum leads to an overdamping of the
modes associated with the order parameter. For the superconductors the
critical behavior is in the $z=2$ universality class. However Pines [32]
showed that the phase diagram of the cuprates superconductors is more
complicated. The normal phase was divided in two regions: one with $z=2$ ($%
\omega _{SF}\sim \xi ^{-2}$) which has no pseudogap and the region with $z=1$
which present a weak pseudogap regime ( $\omega _{SF}\sim \xi ^{-1},\xi
^{-1}=a+bT$ ) and another presenting a strong pseudogap regime ( $\omega
_{SF}\sim \xi ^{-1},\xi ^{-1}=const$ ) separated by a line $T^{*}(x)$ where $%
x$ is the doping. The crossover between $z=1$ and $z=2$ regime was studied
by Sachdev et.al. [33] using the scaling analysis in the framework of $%
\sigma $ model. The main result is that fermions do not neccesary overdamped
and destroy the spin waves. At low damping the crossover is the same as in
pure $\sigma $ mode. The difference is for a finite damping of fermions in
the quantum disorder regime where the energy scale can drive a crossover
from high temperature quantum critical regime $z=1$ to low temperature
quantum critical regime $z=2$.

In this section we will consider for $\chi (\mathbf{q},\omega )$ the
expression:

\begin{equation}
\chi (\mathbf{q},\omega )=\frac{\chi _0(\mathbf{Q})}{\delta +a^2q^2-i\gamma
\omega -\frac{\omega ^2}{\Delta ^2}}
\end{equation}
where $\Delta $ is the gap in the spin wave spectrum.

\vspace{0.1in}

\subsection{Self-Energy of the Electronic Excitations}

\vspace{0.1in}

In the one loop approximation the self-energy of the 2D electronic
excitations interacting with a bosonic mode has the form:

\begin{equation}
\Sigma (\mathbf{p},\omega )=g^2\int \frac{d^2\mathbf{q}}{(2\pi )^2}%
\int_{-\infty }^\infty \frac{d\omega ^{\prime }}{2\pi }Im\chi \left( \mathbf{%
q},\omega ^{\prime }\right) \frac{\coth \frac{\omega ^{\prime }}{2T}-\tanh 
\frac{\varepsilon _{\mathbf{p+q}}}{2T}}{\omega +\omega ^{\prime
}-\varepsilon _{\mathbf{p+q+Q}}+i\delta }
\end{equation}
where $g$ is the coupling constant, $\chi (\mathbf{q},\omega ^{\prime })$
the dynamic susceptibility, $\varepsilon _{\mathbf{p}}=\frac{p^2}{2m}-\mu $
and $\mu $ the chemical potential.

There are two limiting cases for the calculation of the self-energy. The
Fermi-liquid regime appears for $\omega _{SF}\gg T$ and the non-Fermi liquid
regime in the opposite ( renormalized classical ) regime $\omega _{SF}\ll T$
[25]. Perhaps the best known characteristic of a Fermi liquid is that $%
Im\Sigma (\mathbf{k}_F\mathbf{,}\omega ,T=0)\sim \omega ^2$ and $Im\Sigma (%
\mathbf{k}_F\mathbf{,}\omega =0,T)\sim T^2.$

The key to the understanding the Fermi liquid versus non-Fermi liquid regime
is in the relative width in frequency of $\chi ^{\prime \prime }(\mathbf{q}%
,\omega ^{\prime })/\omega ^{\prime }$ versus the width of the combined Bose
and Fermi functions. In general the function $n(\omega ^{\prime })+f(\omega
+\omega ^{\prime })$ ($n(\omega )$ the Bose-Einstein function, $f(\omega )$
the Fermi -Dirac function) depends on $\omega ^{\prime }$ on a scale $%
Max(\omega ,T)$ while far from the phase transition the explicit dependence
of $\chi ^{\prime \prime }(\mathbf{q},\omega ^{\prime })/\omega ^{\prime }$
is on the scale $\omega _{SF}\sim E_F\gg T$. Hence in this case we can
assume that $\chi ^{\prime \prime }(\mathbf{q},\omega ^{\prime })/\omega
^{\prime }$ is constant in the frequency range over which $n(\omega ^{\prime
})+f(\omega +\omega ^{\prime })$ differs from zero. Hence we can approximate
our expression for the imaginary part of the self-energy with:

\[
Im\Sigma (\mathbf{p}_F,\omega )\cong -g^2A(\mathbf{p}_F)\int \frac{d\omega
^{\prime }}\pi \left[ \coth \frac{\omega ^{\prime }}{2T}+\tanh \frac{\omega
+\omega ^{\prime }}{2T}\right] \omega ^{\prime } 
\]

\begin{equation}
=-g^2A(\mathbf{p}_F)\left[ \omega ^2+(\pi T)^2\right]
\end{equation}
where:

\begin{equation}
A(\mathbf{p}_F)=\int \frac{d^2\mathbf{q}}{2\pi }\frac{\chi ^{\prime \prime }(%
\mathbf{q},\omega ^{\prime })}{\omega ^{\prime }}
\end{equation}

As we can see in this regime $Im\Sigma (\omega )\sim \omega ^2$, so the
system behaves as a Fermi liquid. Near an antiferromagnetic phase
transition, the spin fluctuation energy is much smaller than the
temperature. This is the renormalized classical regime. The condition $%
\omega _{SF}\ll T$ means that $\chi ^{\prime \prime }(\mathbf{q},\omega
^{\prime })$ is peaked over a frequency interval $\omega ^{\prime }\ll T$
much narrow than the interval $\omega ^{\prime }\sim T$ over which $n(\omega
^{\prime })+f(\omega +\omega ^{\prime })$ changes. This situation is the
opposite of that encountered in the Fermi liquid regime. To evaluate $%
Im\Sigma $ the Fermi factor can now be neglected compared with classical
limit of the Bose factor $\frac T{\omega ^{\prime }}$. Evaluating the
integral we obtain the exact value obtained using classical fluctuations
[25].

Neglecting the fermionic contribution and using the expansion:

\begin{equation}
\varepsilon _{\mathbf{p+q+Q}}\cong \varepsilon _{\mathbf{p+Q}}+vq\cos \theta
\end{equation}
and Schwinger representation for the propagator:

\begin{equation}
\frac 1i\int_0^\infty dte^{i(\omega -\varepsilon +i\alpha )t}=\frac 1{\omega
-\varepsilon +i\alpha }
\end{equation}
we can write the general form for the self-energy. Let us introduce the
structure factor:

\begin{equation}
S(\mathbf{q},T)=\int_{-\infty }^\infty \frac{d\omega ^{\prime }}{2\pi }%
Im\chi \left( \mathbf{q},\omega ^{\prime }\right) \coth \frac{\omega
^{\prime }}{2T}
\end{equation}
which describe the behavior of the bosonic fluctuations on different energy
scales. Using (2.12) and (2.13) the real and imaginary part $\sum^{\prime
}=Re\sum $ respectively $\sum^{\prime \prime }=Im\sum $ can be calculated
from eq. (2.8) as:

\begin{equation}
\Sigma ^{\prime }(\mathbf{p},\omega )=\frac{g^2}{2\pi }\int_0^\infty dt\sin
\left[ \left( \omega -\varepsilon _{\mathbf{p+Q}}\right) t\right]
\int_0^\infty dqqS(\mathbf{q},T)\int_0^{2\pi }\frac{d\theta }{2\pi }%
e^{-ivt\cos \theta }
\end{equation}

\begin{equation}
\Sigma ^{\prime \prime }(\mathbf{p},\omega )=-\frac{g^2}{2\pi }\int_0^\infty
dt\cos \left[ \left( \omega -\varepsilon _{\mathbf{p+Q}}\right) t\right]
\int_0^\infty dqqS(\mathbf{q},T)\int_0^{2\pi }\frac{d\theta }{2\pi }%
e^{-ivt\cos \theta }
\end{equation}
where $v=\frac pm$. The integral over $\theta $ can be expressed by the
Bessel function:

\begin{equation}
J_0(x)=\frac 1{2\pi }\int_0^{2\pi }d\theta e^{-ix\cos \theta }
\end{equation}
and eqs.(2.14) and (2.15) becomes:

\begin{equation}
\Sigma ^{\prime }(\mathbf{p},\omega )=\frac{g^2}{2\pi }\int_0^\infty dt\sin
\left[ \left( \omega -\varepsilon _{\mathbf{p+Q}}\right) t\right]
\int_0^\infty dqqS(\mathbf{q},T)J_0(qvt)
\end{equation}

\begin{equation}
\Sigma ^{\prime \prime }(\mathbf{p},\omega )=-\frac{g^2}{2\pi }\int_0^\infty
dt\cos \left[ \left( \omega -\varepsilon _{\mathbf{p+Q}}\right) t\right]
\int_0^\infty dqqS(\mathbf{q},T)J_0(qvt)
\end{equation}

These general expressions will be used to calculate the pseudogap in the
electronic energy.

\vspace{0.1in}

\subsection{Pseudogap in the Electronic Energy Spectrum}

\vspace{0.1in}

In this section we calculate the correction to the self-energy due to the
interaction of the electrons with the spin fluctuations for $\omega \ll T$
using eq.(2.7) for dynamic susceptibility.

>From this equation we get:

\begin{equation}
Im\chi (\mathbf{q},\omega )=\frac{\chi _0(\mathbf{Q})\gamma \omega }{\left(
\delta +a^2q^2-\frac{\omega ^2}{\Delta ^2}\right) ^2+\gamma ^2}
\end{equation}
and the structure factor will be approximated as:

\begin{equation}
S(\mathbf{q},T)=\frac{T\gamma \chi _0(\mathbf{Q})}\pi \int_{-\infty }^\infty 
\frac{d\omega }{\left( A^2-\frac{\omega ^2}{\Delta ^2}\right) ^2+\gamma
^2\omega ^2}
\end{equation}
where $A=\delta +a^2q^2$ as we approximated $\coth \frac \omega {2T}\simeq 
\frac{2T}\omega $. The integral in eq. (2.20) can be performed analytically
in the approximation:

\begin{equation}
\frac \Delta {\omega _{SF}}>2\left( \delta +a^2q^2\right)
\end{equation}
as:

\begin{equation}
S(\mathbf{q},T)=\frac{T\chi _0(\mathbf{Q})}{\delta +a^2q^2}
\end{equation}

>From equations (2.17) and (2.18) using the result:

\begin{equation}
\int_0^\infty \frac{J_0(qvt)qdq}{\delta +a^2q^2}=\frac 1{a^2}K_0(tv/\xi )
\end{equation}
and:

\begin{equation}
\int_0^\infty dxK_0(\beta x)\sin (\alpha x)=\frac 1{\sqrt{\alpha ^2+\beta ^2}%
}\ln \left| \frac \alpha \beta +\sqrt{\frac{\alpha ^2}{\beta ^2}+1}\right|
\end{equation}

\begin{equation}
\int_0^\infty dxK_0(\beta x)\cos (\alpha x)=\frac \pi {\sqrt{\alpha ^2+\beta
^2}}
\end{equation}
we obtain the real and imaginary part of the self-energy as:

\begin{equation}
\Sigma ^{\prime }(\mathbf{p},\omega )=\frac{g^2T\chi _0(\mathbf{Q})}{2\pi a}%
\frac 1{\sqrt{\left( \omega -\varepsilon _{\mathbf{p+Q}}\right) ^2+(\frac
v\xi )^2}}\ln \left| \frac{\omega -\varepsilon _{\mathbf{p+Q}}+\sqrt{\left(
\omega -\varepsilon _{\mathbf{p+Q}}\right) ^2+(\frac v\xi )^2}}{\omega
-\varepsilon _{\mathbf{p+Q}}-\sqrt{\left( \omega -\varepsilon _{\mathbf{p+Q}%
}\right) ^2+(\frac v\xi )^2}}\right|
\end{equation}

\begin{equation}
\Sigma ^{\prime \prime }(\mathbf{p},\omega )=-\frac{g^2T\chi _0(\mathbf{Q})}{%
4a^2}\frac 1{\sqrt{\left( \omega -\varepsilon _{\mathbf{p+Q}}\right)
^2+(\frac v\xi )^2}}
\end{equation}

These results are identical with the expressions given by Vilk and Tremblay
[25] using for the dynamic susceptibility a simple form $\chi (\mathbf{q}%
,0)=(\xi ^{-2}+q^2)^{-1}$. The identity of the results is given by the fact
that the details of dynamics are unimportant as long as it does not change
the wave vector dependence.

For the case $\left| \omega -\varepsilon _{\mathbf{p+Q}}\right| >\frac v\xi $
the eqs.(2.26) and (2.27) can be approximated as:

\begin{equation}
\Sigma ^{\prime }(\mathbf{p},\omega )=\frac{g^2T\xi }\delta \frac{\ln
(\left| \omega -\varepsilon _{\mathbf{p+Q}}\right| \xi /v)}{\omega
-\varepsilon _{\mathbf{p+Q}}}
\end{equation}

\begin{equation}
\Sigma ^{\prime \prime }(\mathbf{p},\omega )=-\frac{g^2T\xi }{\omega
-\varepsilon _{\mathbf{p+Q}}}
\end{equation}

At this point we mention that the Green function $G^{-1}(\mathbf{p},\omega
)=G_0^{-1}(\mathbf{p},\omega )-\Sigma (\mathbf{p},\omega )$ will present a
correct behavior ( $G(\omega )\sim 1/\omega $ ) if $\Sigma (\omega )\sim
1/\omega $ if we have:

\begin{equation}
T\ln \frac{\omega _x\xi }v=c
\end{equation}
where $c$ is a constant and $\omega _x$ is the characteristic energy which
is in agreement with the approximations satisfy $\omega _x\ll T$. From
eq.(2.30) we get:

\begin{equation}
\xi (T)=\frac v{\omega _x}\exp \left( \frac cT\right)
\end{equation}
and $\omega _x=\delta U\ll T$, define the proximity to the phase transition.
This simple procedure can be regarded as phenomenology of the Tremblay
renormalized classical regime of the fluctuations obtained in the
self-consistent way, and gives as a main result a temperature independent
for the pseudogap.

However, if we consider the phenomenology proposed by Millis, Monien and
Pines and take $z=1$ the pseudogap obtained from eq. (2.28) as:

\begin{equation}
\Delta _{pg}^2=T\ln \xi
\end{equation}
will present a temperature dependence for the pseudogap.

\vspace{0.1in}

\subsection{Effect of Anisotropy}

\vspace{0.1in}

The experimental data showed a highly anisotropic gap. The effect of the
interplane anisotropy has been considered by Preosti et.al. [34]. Following
the method developed in previous section we will show that there is a
temperature dependence of the pseudogap if the anisotropy is considered. In
this case the structure factor will be taken as:

\begin{equation}
S_a\left( q,q_zT\right) =\frac{T\chi _0(\mathbf{Q})}{\xi ^2}\frac 1{\xi
^{-2}+q^2+\gamma ^2q_z^2}
\end{equation}
where $q^2=q_x^2+q_y^2$, $\gamma =\xi _z/\xi _{xy}$ and $\xi _{xy}=\xi $.
>From eq. (2.17) we obtain:

\begin{equation}
\Sigma ^{\prime }\left( \mathbf{p},\omega \right) =\frac{g^2}{2\pi }%
\int_0^\infty dt\sin \left[ \left( \omega -\varepsilon _{\mathbf{p+Q}%
}\right) t\right] \int_0^{q_c}dqq\int_0^{\overline{q}_c}\frac{dq_z}{q_{BZ}^z}%
S_a\left( q,q_z,T\right) J_0\left( qvt\right)
\end{equation}

In this expression we perform first the integral over variable $t$, which is
different from zero only if:

\begin{equation}
\left| \omega -\varepsilon _{\mathbf{p+Q}}\right| >vq
\end{equation}
and we get the result:

\begin{equation}
\Sigma ^{\prime }\left( \mathbf{p},\omega \right) =\frac{g^2}{2\pi }%
\int_0^{q_c}dqq\int_0^{\overline{q}_c}\frac{dq_z}{q_{BZ}^z}\frac{S_a\left(
q,q_z,T\right) }{\sqrt{\left( \omega -\varepsilon _{\mathbf{p+Q}}\right)
^2-v^2q^2}}
\end{equation}
which will be written using approximation (2.35) as:

\begin{equation}
\Sigma ^{\prime }\left( \mathbf{p},\omega \right) =\frac{\Delta
_{pg}^2\left( T,\gamma \right) }{\left( \omega -\varepsilon _{\mathbf{p+Q}%
}\right) }
\end{equation}
where:

\begin{equation}
\frac{\Delta _{pg}^2\left( T,\gamma \right) }{\Delta _{pg}^2\left( 0\right) }%
\frac{T_0}T=\frac{\overline{q}_c}{q_{BZ}^z}A\left( \xi ,\gamma \right)
\end{equation}

\begin{figure}[tbh]
\centering
\includegraphics[clip,width=0.8\textwidth]{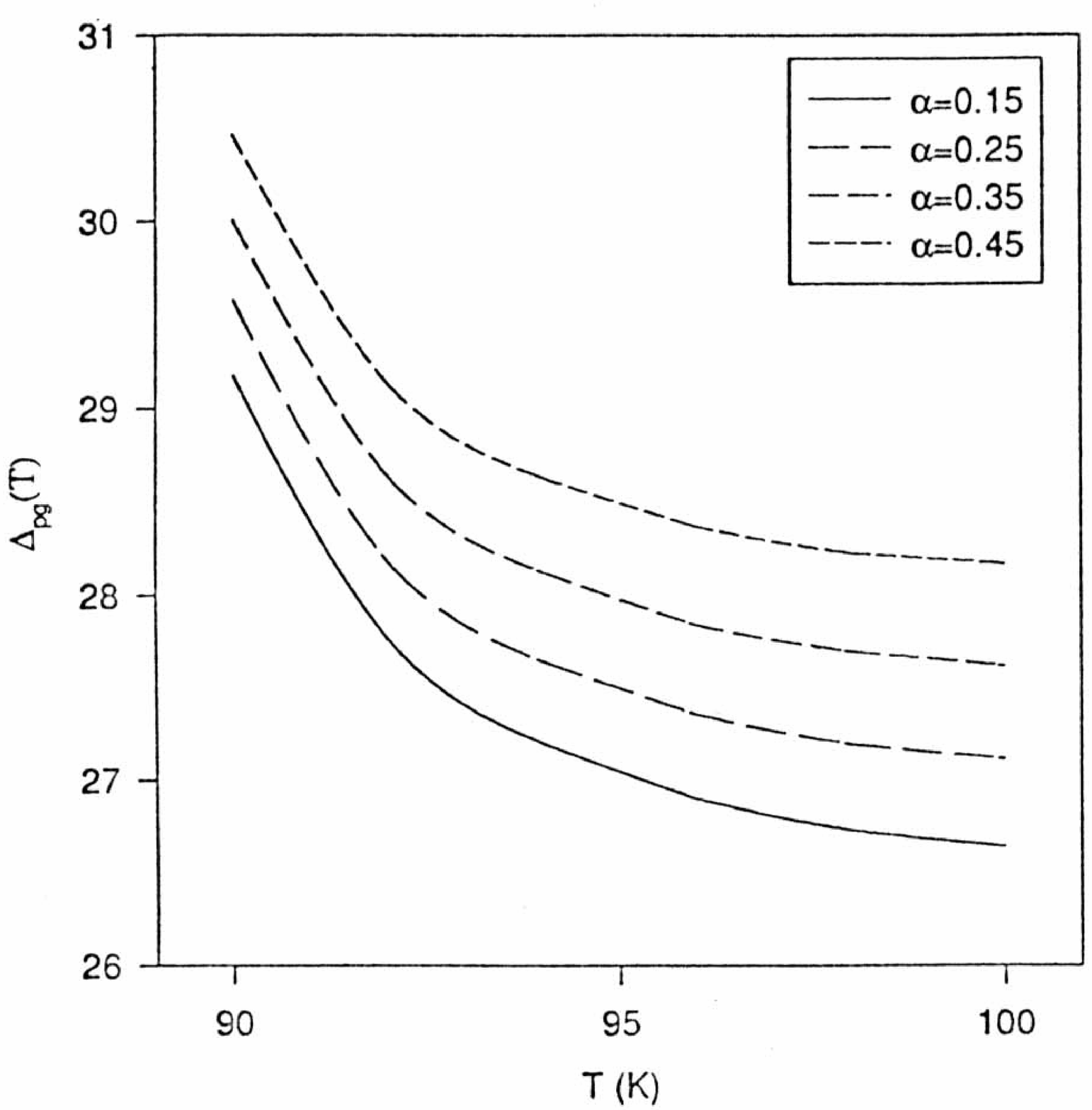}
\caption{Temperature dependence of the
pseudogap $\Delta (T)$ for $\gamma =0.9$ and $\xi ^{-1}(T)=\exp [100T_0/T]$}
\label{Fig2}
\end{figure}
%\FRAME{ftbpFU}{3.7792in}{2.8089in}{0pt}{\Qcb{}%
%}{}{fig3.jpg}{\special{language "Scientific Word";type "GRAPHIC";display
%"PICT";valid_file "F";width 3.7792in;height 2.8089in;depth 0pt;cropleft
%"-0";croptop "1.0106";cropright "0.9988";cropbottom "0.0114";filename
%'D:/TEZA/FIG3.JPG';file-properties "XNPEU";}}with:
%
\begin{equation}
T_0=\frac{2\pi \xi ^2}{\chi _0(\mathbf{Q})}\Delta _{pg}(0)
\end{equation}
and:

\begin{equation}
A\left( \xi ,\gamma \right) =\frac{\sqrt{1+\left( \xi \overline{q}_c\right)
^2}}\gamma \arctan \frac \gamma {\sqrt{1+\left( \xi \overline{q}_c\right) ^2}%
}-\frac{\arctan \gamma \xi \overline{q}_c}{\gamma \xi \overline{q}_c}+\frac
12\ln \frac{1+\left( \gamma \xi \overline{q}_c\right) ^2}{1-\left( \gamma
\xi \overline{q}_c\right) ^2}
\end{equation}

We can take a soft cutoff taking $\chi _0(\mathbf{Q})\rightarrow \chi _0(%
\mathbf{Q})[1-q_c^{-2}(q^2+\gamma q_z^2)]$ and the equation for the
pseudogap has the form:

\begin{equation}
\frac{\Delta _{pg}^2\left( T,\gamma \right) }{\Delta _{pg}^2(0)}\frac{T_0}T=%
\frac{\overline{q}_c}{q_{BZ}^z}\left[ \left( 1+\left( \xi \overline{q}%
_c\right) ^2\right) A\left( \xi ,\gamma \right) -\frac 12\right]
\end{equation}

\begin{figure}[tbh]
\centering
\includegraphics[clip,width=0.8\textwidth]{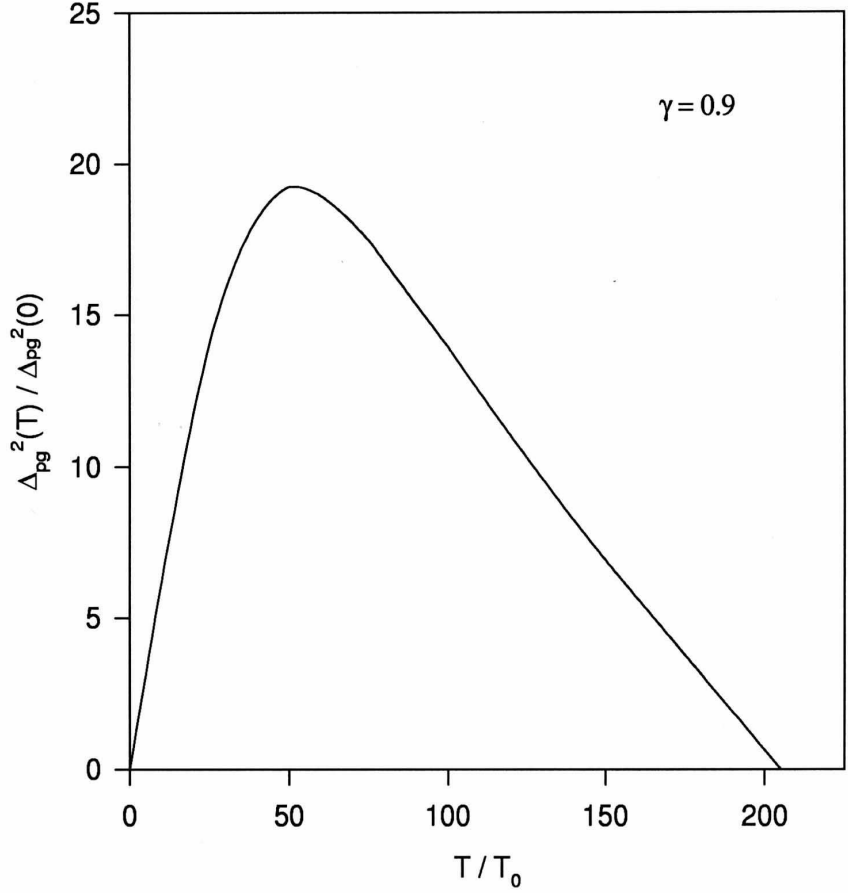}
\caption{Temperature dependence of the
pseudogap $\Delta (T)$ for $\xi ^{-1}(T)=10^{-3}+10^{-2}T/T_0$,$\gamma =0.9$
respectively $\gamma =1$.}
\label{Fig3}
\end{figure}
%\FRAME{ftbpFU}{3.7498in}{2.7285in}{0pt}{\Qcb{}}{}{fig4.jpg}
%{\special{language "Scientific
%Word";type "GRAPHIC";display "PICT";valid_file "F";width 3.7498in;height
%2.7285in;depth 0pt;cropleft "0.0082";croptop "1.0007";cropright
%"1.0073";cropbottom "0";filename 'D:/TEZA/FIG4.JPG';file-properties "XNPEU";}%
%}

If we define the temperature $T^{*}(\gamma )$ by:

\begin{equation}
\xi (T^{*}(\gamma ))\rightarrow \infty
\end{equation}
we can show that:

\begin{equation}
\lim_{\gamma \rightarrow 0}T^{*}(\gamma )=0
\end{equation}

We have to mention that there is an important difference between this result
and the result from Ref.[34]. In this model $\xi $ is the correlation length
of the magnetic fluctuations, differently from Ref.[34] where $\xi $ is the
superconducting correlation length. The condition (2.42) is equivalent to
the proximity of the magnetic quantum phase transition. Even if this model
is realistic taking into consideration the magnetic fluctuations the
matching between the pseudogap and superconducting gap was not solved. An
important point of the model is the temperature dependence of the magnetic
correlation length. In Fig.2.1 we present the temperature dependence of the
pseudogap obtained from eq. (27) and $\xi ^{-1}(T)=10^{-3}+10^{-2}T$. Using $%
\gamma =0.9$ and $\xi ^{-1}=10^{-2}+10^{-3}T/T_0$ we obtained a dependence
of the pseudogap given by eq. (2.38) in Fig. 2.2.

\vspace{0.1in}

\section{Electron-Fluctuation Interaction in a non-Fermi Liquid
Superconductor}

\vspace{0.1in}

In this section we consider the model introduced in [1] and describe in the
first section of this chapter. We consider that the superconducting state
appears due to an attractive interaction and is describe by the BCS-like
order parameter $\Delta _k$ which can be calculated from the Gorkov
equations. The fluctuations of this parameter can interact with the
electrons and the fermionic spectrum of the elementary excitations changes.
Such an effect has been studied in the BCS superconductors by different
authors [25], [35] and it was showed that this interaction gives a
contribution to the density of states for $T>T_c$ which explained the
behavior of the tunnelling measurements.

For a superconductor described by the Gorkov-like equations with the normal
state described by eq. (2.1) the propagator of the fluctuations has the
expression:

\begin{equation}
D\left( \mathbf{q},i\omega _n\right) =\frac 1{V^{-1}+\Pi \left( \mathbf{q}%
,i\omega _n\right) }
\end{equation}
where $V$ is the attractive interaction between the electrons and $\Pi
\left( \mathbf{q},i\omega _n\right) $ is the polarization operator defined
as:

\begin{equation}
\Pi \left( \mathbf{q},i\omega _n\right) =T\sum_{\omega _l}\int \frac{d^2%
\mathbf{p}}{(2\pi )^2}G\left( \mathbf{p},i\omega _l\right) G\left( \mathbf{%
q-p},i\omega _n-i\omega _l\right)
\end{equation}
where $G\left( \mathbf{p},i\omega _l\right) $ is the Green function related
to electrons, which in terms of Dyson equation has the following for:

\begin{equation}
G^{-1}\left( \mathbf{p},i\omega _l\right) =G_0^{-1}\left( \mathbf{p},i\omega
_l\right) -\Sigma \left( \mathbf{p},i\omega _l\right)
\end{equation}
where the self-energy is given by:

\begin{equation}
\Sigma \left( \mathbf{p},i\omega _l\right) =-T\sum_{\omega _l}\int \frac{d^2%
\mathbf{q}}{(2\pi )^2}D\left( \mathbf{q},i\omega _n\right) G\left( \mathbf{%
q-p},i\omega _n-i\omega _l\right)
\end{equation}

Equations (2.44)-(2.47) has to be solved self-consistently, but this cannot
be done analytically. However, in the mode coupling approximation it can be
done and we can calculate the new energy of the electronic excitations. We
mention that in eq. (2.47) the vertex corrections have been neglected in
order to developed a simple analytical model.

\vspace{0.1in}

\subsection{Mode Coupling Approximation}

\vspace{0.1in}

In this approximation we consider first that $G\left( \mathbf{p},i\omega
_n\right) \approx G_0\left( \mathbf{p},i\omega _n\right) $ and from eq.
(2.45) we define the polarization

\begin{equation}
\Pi _0\left( \mathbf{q},i\omega _m\right) =\int \frac{d^2\mathbf{k}}{(2\pi
)^2}S\left( \mathbf{k,q},i\omega _m\right)
\end{equation}
where:

\begin{equation}
S\left( \mathbf{k,q},i\omega _m\right) =(-1)^{1-\alpha }T\sum_{\omega _n}%
\frac{g^2(\alpha )\omega _c^{-2\alpha }}{(i\omega _n-\varepsilon _{\mathbf{k}%
})^{1-\alpha }(i\omega _n-i\omega _m-\varepsilon _{\mathbf{q-k}})^{1-\alpha }%
}
\end{equation}

We performed the analytical calculation of $\Pi _0(\mathbf{q},i\omega _m)$
given by eq. (2.48) and from the eq. (2.44) the propagator for the order
parameter fluctuations has been obtained as:

\[
D_0^{-1}\left( \mathbf{q},i\omega _n\right) =N(0)A(\alpha )\left\{ C(\alpha
)\left[ \left( \frac T{\omega _c}\right) ^{2\alpha }-\left( \frac{T_c}{%
\omega _c}\right) ^{2\alpha }\right] \right. 
\]

\[
+\frac{i\omega _n(1-\alpha )}TM\left( \alpha ,\frac T{\omega _c},\frac{%
\omega _D}{\omega _c}\right) 
\]

\begin{equation}
+\left. \left( \frac{v_Fq}{2T}\right) ^2(1-\alpha )^2N\left( \alpha ,\frac
T{\omega _c},\frac{\omega _D}{\omega _c}\right) \right\}
\end{equation}
where the critical temperature has been obtained as [7]:

\begin{equation}
T_c^{2\alpha }=\frac 1{C(\alpha )}\left[ D(\alpha )\omega _D^{2\alpha }-%
\frac{\omega _c^{2\alpha }}{A(\alpha )N(0)V}\right]
\end{equation}
and the constants from eqs.(2.50) and (2.51) are:

\begin{equation}
A(\alpha )=g^2(\alpha )\frac{2^{2\alpha }}\pi \sin \pi (1-\alpha )
\end{equation}

\begin{equation}
C(\alpha )=\Gamma ^2(\alpha )[1-2^{1-2\alpha }]\zeta (2\alpha )
\end{equation}

\begin{equation}
D(\alpha )=\frac{\Gamma (1-2\alpha )\Gamma (\alpha )}{2\alpha \Gamma
(1-\alpha )}
\end{equation}

\[
M\left( \alpha ,\frac T{\omega _c},\frac{\omega _D}{\omega _c}\right) =\frac{%
\Gamma (\alpha -1)\Gamma (\alpha -1/2)}{2\sqrt{\pi }}\left[ 1-2^{2-2\alpha
}\right] \zeta (2\alpha -1)\left( \frac T{\omega _c}\right) ^{2\alpha } 
\]

\begin{equation}
-\frac{B(2-2\alpha ,\alpha -1)}{2(2\alpha -1)}\left( \frac{\omega _D}{\omega
_c}\right) ^{2\alpha -1}\left( \frac T{\omega _c}\right)
\end{equation}
and:

\[
N\left( \alpha ,\frac T{\omega _c},\frac{\omega _D}{\omega _c}\right)
=\left[ \frac{2\Gamma (\alpha -2)\Gamma (\alpha -1/2)}{\sqrt{\pi }}+\Gamma
^2(\alpha -1)\right] 
\]

\[
\times \frac{1-2^{3-2\alpha }}4\zeta (2-2\alpha )\left( \frac T{\omega
_c}\right) ^{2\alpha } 
\]

\begin{equation}
-\frac{B(3-2\alpha ,\alpha -2)}{4(2\alpha -2)}\left( \frac{\omega _D}{\omega
_c}\right) ^{2\alpha -2}\left( \frac T{\omega _c}\right) ^2
\end{equation}

$B(x,y)=\Gamma (x)\Gamma (y)/\Gamma (x+y)$ and $\Gamma (x)$ is the Euler
function and $\zeta (x)$ is the Riemann function. Using a similar form with
one introduced by Schmidt the fluctuation propagator will be written as:

\begin{equation}
D_0^{-1}\left( \mathbf{q},i\omega _n\right) =N(0)\left[ b(\alpha )\tau
(\alpha )+ia(\alpha )\omega _n+\xi ^2(\alpha ,T)q^2\right]
\end{equation}
where:

\begin{equation}
\tau (\alpha )=\left( \frac T{T_c}\right) ^{2\alpha }-1
\end{equation}

\begin{equation}
a(\alpha )=\frac{M\left( \alpha ,\frac T{\omega _c},\frac{\omega _D}{\omega
_c}\right) }T(1-\alpha )A(\alpha )
\end{equation}

\begin{equation}
b(\alpha )=A(\alpha )C(\alpha )\left( \frac T{\omega _c}\right) ^{2\alpha }
\end{equation}
and:

\begin{equation}
\xi (\alpha )=\frac{v_F^2(1-\alpha )^2}{4T^2}N\left( \alpha ,\frac T{\omega
_c},\frac{\omega _D}{\omega _c}\right) A(\alpha )
\end{equation}

In the approximation $\Sigma \ll \pi T$ the Green function given by eq.
(2.46) will be approximated as $G=G_0+G_0\Sigma G_0$ and $\Pi $ will be
modified by $\delta \Pi $ also linear in $\Sigma $. Following Ref.[11] we
calculated $\delta \Pi $ in the ''box approximation'' as:

\begin{equation}
\delta \Pi =2T^2\sum_{\omega _n}\int \frac{d^2\mathbf{p}}{(2\pi )^2}%
G_0^2\left( \mathbf{p},i\omega _n\right) G^2\left( -\mathbf{p},-i\omega
_n\right) \int \frac{d^2\mathbf{q}}{(2\pi )^2}D\left( \mathbf{q},\omega
_n=0\right)
\end{equation}
where:

\begin{equation}
D_0^{-1}\left( \mathbf{q},i\omega _n\right) =V^{-1}+\Pi \left( \mathbf{q}%
,i\omega _n\right) +\delta \Pi \left( \mathbf{q},i\omega _n\right)
\end{equation}

The box vertex describes the interaction between fluctuations and becomes
important only in the critical regime in the standard superconductors. In
order to calculate $\delta \Pi (\mathbf{q},i\omega _n)$ we introduce:

\begin{equation}
B_0=\frac 1{N(0)}T\sum_{\omega _n}\int \frac{d^2\mathbf{p}}{(2\pi )^2}%
G_0^2\left( \mathbf{p},i\omega _n\right) G_0^2\left( -\mathbf{p},-i\omega
_n\right)
\end{equation}
where $N(0)=m/2\pi $ is the density of states. If we used for the electronic
Green function eq.(2.1) we obtain:

\begin{equation}
B_0(T)=\frac{B\left( 1/2,3/2-2\alpha \right) }\pi \frac{\left[ 2^{3-4\alpha
}-1\right] \zeta (3-4\alpha )}{2^{3-4\alpha }}\frac{\omega _c^{-4\alpha }}{%
(\pi T)^{2-4\alpha }}g^4(\alpha )
\end{equation}

If we introduce $\widetilde{\tau }(\alpha )=\tau (\alpha )+\delta \Pi /N(0)$
the fluctuation propagator given by eq. (2.57) will be:

\begin{equation}
D^{-1}\left( \mathbf{q},i\omega _n\right) =b(\alpha )\widetilde{\tau }%
(\alpha )+ia(\alpha )\omega _n+\xi ^2\left( \alpha ,T\right) q^2
\end{equation}
where:

\begin{equation}
\widetilde{\tau }(\alpha )-\tau (\alpha )=\frac{2B_0(T)}{N(0)}T\int \frac{d^2%
\mathbf{q}}{(2\pi )^2}\frac 1{\widetilde{\tau }(\alpha )+\xi ^2(\alpha
,T)q^2}
\end{equation}

If we perform this integral taking the upper limit $q_M=1/\xi (\alpha ,T)$
from eq.(2.64) we get:

\begin{equation}
\widetilde{\tau }(\alpha )-\tau (\alpha )=\frac{B_0(T)T}{2\pi N(0)\xi
^2(\alpha ,T)}\ln \frac{1+\widetilde{\tau }(\alpha )}{\xi ^2(\alpha ,T)}
\end{equation}

For realistic parameters ( $T_c=100K,\omega _c=200K$ ) the deference $%
\widetilde{\tau }(\alpha )-\tau (\alpha )$ becomes important only near a
critical value of $\alpha $ defined by $\xi (\alpha _c)=0$. In the BCS limit
( $\alpha =0$ ) this parameter is small and this behavior can be associated
with the occurrence of the preformed pairs in the domain $T_c<T<T^{*}$,
controlled by $\alpha $. This behavior is in fact due to the occurrence of a
pseudogap in the electronic excitations.

\vspace{0.1in}

\subsection{Electronic Self-Energy}

\vspace{0.1in}

The self energy due to the interaction between electrons and fluctuations is
given by eq.(2.47) where $D(\mathbf{q},i\omega _n)$ is given by eq.(2.66).
First we calculate the summation over Matsubara frequencies $\omega _n$ :

\begin{equation}
S=T\sum_{\omega _l}D\left( \mathbf{q},i\omega _n\right) G\left( \mathbf{q-p}%
,i\omega _n-i\omega _l\right)
\end{equation}

\begin{equation}
S=T\sum_{\omega _n}\frac{(-1)^\alpha g(\alpha )\omega _c^{-\alpha }e^{i\pi
\alpha /2}}{N(0)\left( b\widetilde{\tau }+ia\omega _n+\xi ^2q^2\right)
\left( i\omega _l-i\omega _n+\varepsilon _{\mathbf{k}}\right) ^{1-\alpha }}
\end{equation}
transforming this sum in a contour integral which has a pole at $\Omega (%
\mathbf{q})=-(b\widetilde{\tau }+\xi ^2q^2)/a$ and a cut line from $i\omega
_l+\varepsilon _{\mathbf{k}}$ to $\infty $ in the upper semiplane. From eq.
(2.59) we can see that $a(\alpha )=-\left| a(\alpha )\right| $ and $\Omega (%
\mathbf{q})=-(b\widetilde{\tau }+\xi ^2q^2)/\left| a\right| $. Performing
this integral we obtain:

\begin{equation}
S=\frac{\omega _c^{-\alpha }}{N(0)}\frac{n(\Omega (\mathbf{q}))g(\alpha
)e^{i\pi \alpha /2}}{[-i\omega _l-\varepsilon _{\mathbf{k}}-\Omega (\mathbf{q%
})]^{1-\alpha }}-\frac{\omega _c^{-\alpha }}{N(0)}\frac{\sin [\pi (1-\alpha
)]}\pi
\end{equation}

\begin{equation}
\times \int_{\varepsilon _{\mathbf{k}}}^\infty dt\frac{f(t)g(\alpha )e^{i\pi
\alpha /2}}{[b\widetilde{\tau }-\left| a\right| (t+i\omega _l)+\xi
^2q^2](t-\varepsilon _{\mathbf{k}})^{1-\alpha }}
\end{equation}
where $n(x)$ is the Bose-Einstein function and $f(x)$ is the Fermi-Dirac
function and \thinspace $\varepsilon _{\mathbf{k}}=k^2/2m-E_F$. The integral
from the second contribution in eq.(2.72) will be performed using the
expansion:

\begin{equation}
f(t)=\sum_{m=0}^\infty (-1)^m\exp \left( -\beta (m+1)t\right)
\end{equation}
and the last term become:

\[
I_1=\sum_{m=0}^\infty \frac{(-1)^m}{\left| a\right| }\frac{[\varepsilon _{%
\mathbf{k}}+\Omega (\mathbf{q})]^{\alpha /2+1}}{[\beta (m+1)]^{\alpha /2}}%
\exp \left( \frac{\beta (m+1)[\Omega (\mathbf{q})-\varepsilon _{\mathbf{k}}]}%
2\right) 
\]

\begin{equation}
\times \Gamma (\alpha )W_{-\alpha /2,\alpha /2-1/2}\{\beta (m+1)[\varepsilon
_{\mathbf{k}}+\Omega (\mathbf{q})]\}
\end{equation}
where the Whittaker function $W_{\lambda ,\mu }(z)$ will be approximated as:

\begin{equation}
W_{\lambda ,\mu }(z)\cong e^{-z/2}z^\lambda
\end{equation}

These results gives for eq.(2.72) the expression:

\[
S=\frac{\omega _c^{-\alpha }}{N(0)}\frac{n(\Omega (\mathbf{q}))g(\alpha
)e^{i\pi \alpha /2}}{[-i\omega _l-\varepsilon _{\mathbf{k}}-\Omega (\mathbf{q%
})]^{1-\alpha }}+g(\alpha )e^{i\pi \alpha /2}\frac{\omega _c^{-\alpha }}{N(0)%
} 
\]

\[
\times \frac{\sin [\pi (1-\alpha )]}\pi \sum_{m=0}^\infty \frac{(-1)^m}{%
\left| a\right| }\frac{\varepsilon _{\mathbf{k}}+i\omega _l+\Omega (\mathbf{q%
})}{[\beta (m+1)]^\alpha } 
\]

\begin{equation}
\times \Gamma (\alpha )\exp [\beta (m+1)\varepsilon _{\mathbf{k}}]
\end{equation}

In the limit $k\cong k_F$ the second term denoted as $S_2$ becomes:

\[
S_2=\frac{\omega _c^{-\alpha }}{N(0)}\frac{\sin [\pi (1-\alpha )]}\pi \frac{%
i\omega _l\left| a\right| +b\widetilde{\tau }+\xi ^2q^2}{\left| a\right| ^2} 
\]

\begin{equation}
\times \Gamma (\alpha )(1-2^{1-\alpha })\zeta (\alpha )g(\alpha )e^{i\pi
\alpha /2}
\end{equation}
and if $T\rightarrow T_c,\omega _l\rightarrow 0,q\rightarrow 0$ this term
can be neglected. This approximation is in fact equivalent with the physical
picture proposed by Vilk and Tremblay [25] in which the occurrence of the
pseudogap is given by the interaction between the electrons and the
classical fluctuations. Indeed, in this regime the first term of eq.(2.76)
can be written as:

\begin{figure}[tbh]
\centering
\includegraphics[clip,width=0.8\textwidth]{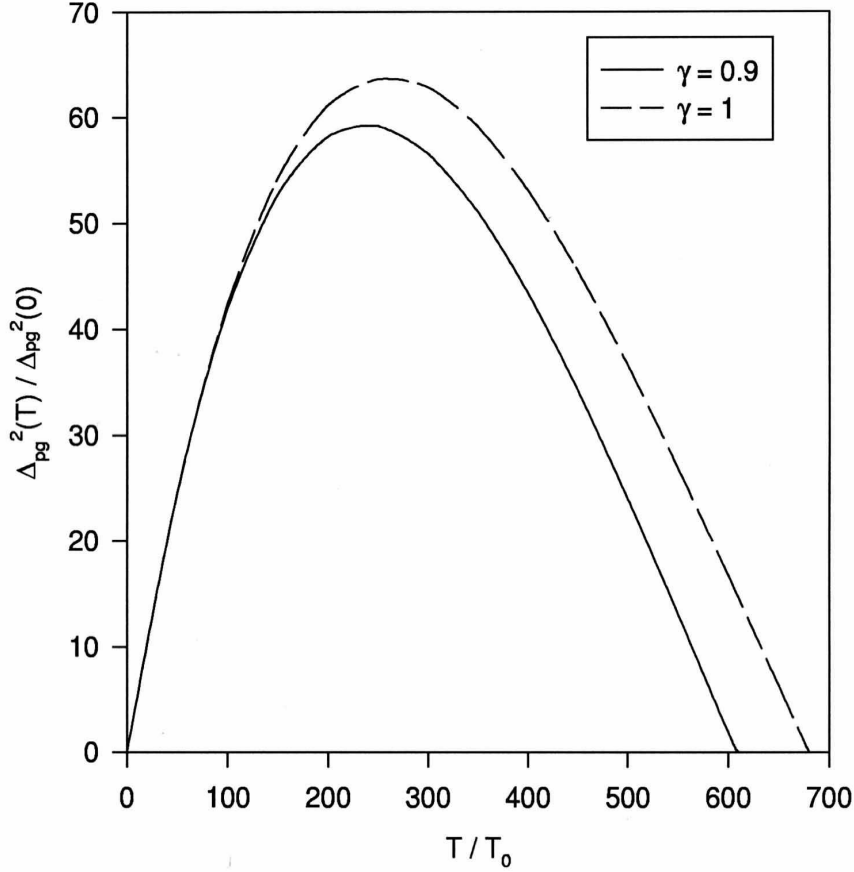}
\caption{The temperature dependence for
the pseudogap for $\xi _{q_M}\sim k_F\xi \sim 10,T_c=100$}
\label{Fig4}
\end{figure}
%\FRAME{ftbpFU}{3.9193in}{2.6489in}{0pt}{\Qcb{}}{}{fig2.jpg}{%
%\special{language "Scientific Word";type "GRAPHIC";display "PICT";valid_file
%"F";width 3.9193in;height 2.6489in;depth 0pt;cropleft "0";croptop
%"1.0013";cropright "1.0007";cropbottom "0";filename
%'D:/TEZA/FIG2.JPG';file-properties "XNPEU";}}
%
\begin{equation}
S\cong \frac 1{N(0)}n[\Omega (\mathbf{q})]G\left( \mathbf{k},-i\omega
_l+\Omega (\mathbf{q})\right)
\end{equation}
and the electronic self-energy becomes:

\begin{equation}
\Sigma \left( \mathbf{p},\omega +i0\right) \cong -\Delta _{pg}^2G\left( 
\mathbf{k},-i\omega _l\right)
\end{equation}
where we considered $\varepsilon _{\mathbf{k}}\gg $ $\Omega (\mathbf{q})$
and:

\begin{equation}
\Delta _{pg}^2=\frac 1{N(0)\left| a\right| }\int \frac{d^2\mathbf{q}}{(2\pi
)^2}n[\Omega (\mathbf{q})]
\end{equation}
will be approximated as:

\begin{equation}
\Delta _{pg}^2=\frac T{2\pi N(0)\left| a\right| }\int_0^{q_M}\frac{qdq}{(b%
\widetilde{\tau }+\xi ^2q^2)/\left| a\right| }
\end{equation}
where $q_M$ is the wave number cutoff. From eq. (2.81) we calculate the
temperature dependence of $\Delta _{pg}^2(T)$ as:

\begin{equation}
\Delta _{pg}^2=\frac T{4\pi N(0)\left| a\right| }\ln \left( 1+\frac{\xi ^2}{b%
\widetilde{\tau }}q_M^2\right)
\end{equation}

The temperature dependence of $\Delta _{pg}$ given by eq.(2.81) using the
approximations $\widetilde{\tau }(\alpha )\cong \tau (\alpha )$ and $\ln (1+%
\frac{\xi ^2}{b\widetilde{\tau }}q_M^2)\cong \ln (\frac{\xi ^2}{b\widetilde{%
\tau }}q_M^2)$ is shown in Fig.2.3 for $\xi q_M\sim k_F\xi \sim 10,4\pi
N(0)\xi \sim 1$, and $T_c=89K$.

\vspace{0.1in}

\section{Field-Theoretical Description of Crossover Between BCS and BEC
in a non-Fermi Superconductor}

\vspace{0.1in}

The problem of the crossover from BCS superconducting state to a
Bose-Einstein condensate (BEC) of local pairs [45-46] becomes very important
in the context of high temperature superconductors (HTSC). While at the
present time there is no quantitative microscopic theory for the occurrence
of the superconducting state in the doped antiferromagnetic materials, it is
generally accepted that the superconducting state can be described in terms
of a pairing picture. The short coherence length ($\xi \sim $ 10-20 A)
increased the interest for the problem [48-58] because it showed that the
BCS equations of highly overlapping pairs, or the description in terms of
composite bosons cannot describe the whole regime between weak and strong
coupling. The mean field method developed by different authors [47,48,50,58]
and solved analytically in two and three dimension, and the Ginzburg-Landau
description [51,52,57] showed that the evolution between the two limits is
continues, no singularities during this evolution appearing. The zero
temperature coherence length in the framework of field-theoretical method
has been studied in Ref. [56]. The problem of the BCS-BEC crossover in
arbitrary dimension $d$ using the field-theoretical method has been
extensively discussed in Refs. [58-61], where the chemical potential, the
number of condensed pairs and the repulsive interaction between pairs have
been calculating using the analogy with the field-theoretical description of
superfluidity. In this section we apply this method to study the crossover
problem for a non-Fermi superconductor described by the Anderson model [1] .

\vspace{0.1in}

\subsection{Weak Coupling Limit}

\vspace{0.1in}

The BCS-like model for the non-Fermi system is described by the Lagrangian

\begin{equation}
L=\psi _{\uparrow }^{+}G^{-1}\psi _{\uparrow }+\psi _{\downarrow
}^{+}(G^{-1})^{*}\psi _{\downarrow }-\lambda _0\psi _{\uparrow }^{+}\psi
_{\downarrow }^{+}\psi _{\downarrow }\psi _{\uparrow }
\end{equation}
where the normal state is describe by the Green function (2.1)

If we introduce the two-component fermionic field

\begin{equation}
\Psi =\left( 
\begin{array}{c}
\psi _{\uparrow } \\ 
\psi _{\downarrow }^{+}
\end{array}
\right) \hspace{1.0in}\Psi ^{+}=\left( \psi _{\uparrow }^{+}\psi
_{\downarrow }\right)
\end{equation}
the non-interacting part of the Lagrangian (1) is:

\begin{equation}
L_0=\Psi ^{+}\left( 
\begin{array}{cc}
G^{-1} & 0 \\ 
0 & G^{-1}
\end{array}
\right) \Psi
\end{equation}

In order to calculate the partition function:

\begin{equation}
Z=\int D\Psi ^{+}D\Psi \exp \left( i\int_xL\right)
\end{equation}
we will transform the interaction contribution from the Lagrangian (2.83) as:

\begin{equation}
\exp \left( -i\lambda _0\int_x\psi _{\uparrow }^{+}\psi _{\downarrow
}^{+}\psi _{\downarrow }\psi _{\uparrow }\right) =\int D\Delta ^{+}D\Delta
\exp \left( -i\int_x\left( \Delta ^{+}\psi _{\downarrow }\psi _{\uparrow
}+\psi _{\uparrow }^{+}\psi _{\downarrow }^{+}\Delta -\frac 1{\lambda
_0}\Delta ^{+}\Delta \right) \right)
\end{equation}
where $\int_x=\int dt\int d^dx$ as in Ref.[58-61] and $\Delta =\lambda
_0\psi _{\downarrow }\psi _{\uparrow }$ is a bosonic field. The partition
function defined by eq. (2.86) will be expressed using eq. (2.87) in a
bilinear form as

\begin{equation}
Z=\int D\Psi ^{+}D\Psi \int D\Delta ^{+}D\Delta \exp \left( \frac i{\lambda
_0}\Delta ^{+}\Delta \right) \exp \left[ i\int_x\Psi ^{+}\left( 
\begin{array}{cc}
G^{-1} & \Delta \\ 
\Delta & G^{-1}
\end{array}
\right) \Psi \right]
\end{equation}

Performing the integral over the Grassmann fields the partition function
becomes

\begin{equation}
Z=\int D\Delta ^{+}D\Delta \exp \left( iS_{eff}[\Delta ^{+},\Delta ]+\frac
1{\lambda _0}\Delta ^{+}\Delta \right)
\end{equation}
where $S_{eff}[\Delta ^{+},\Delta ]$ is the one loop effective action,which
can be written as:

\begin{equation}
S_{eff}[\Delta ^{+},\Delta ]=-iTr\ln \left( 
\begin{array}{cc}
f(\alpha )(p_0-\varepsilon _{\mathbf{p}})^{1-\alpha } & -\Delta \\ 
-\Delta ^{+} & f(\alpha )(p_0+\varepsilon _{\mathbf{p}})^{1-\alpha }
\end{array}
\right)
\end{equation}
where $f(\alpha )=\omega _c^\alpha g^{-1}(\alpha )$ and the $Tr$ has been
used according to the meaning from Ref. [60].

In the mean field approximation the integral from eq. (2.89) can be
performed using the solution given by the saddle point and for $T\neq 0$ the
critical temperature $T_c$ is given by the eq. (2.51) replacing $V$ with $%
\lambda _0$ . This expression is valid only in the limit $0<\alpha <0.5$ and
a positive critical temperature implies for the coupling constant the
condition $\left| \lambda _0\right| >\lambda _c$, with $\lambda _c=(\omega
_c/\omega _D)^\alpha /(A(\alpha )/D(\alpha ))$. We have to mention that the
critical temperature obtained is different from one obtained in Ref. [15]
and it is easy to show that it gives the exact BCS result in the limit $%
\alpha \rightarrow 0$. If we consider the effective action as:

\[
S_{eff}[\Delta ^{+},\Delta ]=-iTr\ln \left( 
\begin{array}{cc}
f(\alpha )(p_0-\varepsilon _{\mathbf{p}})^{1-\alpha } & 0 \\ 
0 & f(\alpha )(p_0+\varepsilon _{\mathbf{p}})^{1-\alpha }
\end{array}
\right) 
\]

\begin{equation}
-iTr\ln \left( 1-\frac{\left| \overline{\Delta }\right| }{f^2(\alpha
)(p_0^2-\varepsilon _{\mathbf{p}}^2)^{1-\alpha }}\right)
\end{equation}
and the system as space-time independent the partition function can be
written as:

\begin{equation}
Z=Z_0\exp \left( \frac i{\lambda _0}\overline{\Delta }^{+}\overline{\Delta }%
\right)
\end{equation}
$Z_0$ containing the non-interacting contribution, and we get for the
renormalization coupling constant $\lambda $ the expression:

\begin{equation}
\frac 1\lambda =\frac 1{\lambda _0}+\frac i{f^2(\alpha )}\int \frac{d^2%
\mathbf{p}}{(2\pi )^2}\int \frac{dp_0}{2\pi }\frac 1{\left(
p_0^2-\varepsilon _{\mathbf{p}}^2\right) ^{1-\alpha }}
\end{equation}

Using the integral:

\begin{equation}
\int_{k_0}\frac 1{\left( k_0^2-E^2+i\eta \right) ^l}=i(-1)^l\sqrt{\pi }\frac{%
\Gamma (l-1/2)}{\Gamma (l)}\frac 1{E^{2l-1}}
\end{equation}
we calculate $\lambda $ as

\begin{equation}
\frac 1\lambda =\frac 1{\lambda _0}+\frac 1{\lambda _1}
\end{equation}
where

\begin{equation}
\lambda _1=-\frac{4\pi \alpha g^{-2}(\alpha )}{\cos (\pi (\alpha -1))}\frac
1{B(1/2,1/2-\alpha )}\left( \frac{\omega _c}{\omega _D}\right) ^{2\alpha }
\end{equation}

The expression given by eq. (2.96) is positive for $\alpha <0.5$. The new
coupling constant $\lambda $ has to be also negative in order to have
superconductivity ( $\lambda <0$ ) and this condition is satisfied if $%
\left| \lambda _0\right| <\lambda _1$. If we consider also the condition $%
\left| \lambda _0\right| >\lambda _c$ we get the general condition for the
coupling constant $\lambda _0$, $\lambda _c<\left| \lambda _0\right|
<\lambda _1$, which is satisfied for $0<\alpha <0.5$.

We mention that for the weak coupling limit $\lambda _0\rightarrow 0^{-}$
the BCS limit studied in Ref. [15] is reobtained, but we also showed that
the critical constant calculated from the critical temperature is $\lambda
_1(\alpha \rightarrow 0)=0.$

In the limit $\lambda _0\rightarrow -\infty $ called the strong coupling
limit, we expect an important effect of the non-Fermi character of the
electrons in the coupling constant.

\vspace{0.1in}

\subsection{Strong Coupling Limit}

\vspace{0.1in}

In this limit we consider $\Delta (x)=\overline{\Delta }+\widetilde{\Delta }%
(x)$ and consider the action $S_{eff}[\widetilde{\Delta }^{+},\widetilde{%
\Delta }]$ obtained from eq. (2.90) as:

\begin{equation}
S_{eff}\left( \widetilde{\Delta }^{+},\widetilde{\Delta }\right) =-iTr\left(
1+\widehat{G}\widehat{\widetilde{\Delta }}\right)
\end{equation}
where:

\begin{equation}
\widehat{G}^{-1}=\left( 
\begin{array}{cc}
f(\alpha )(p_0-\varepsilon _{\mathbf{p}})^{1-\alpha } & -\overline{\Delta }
\\ 
-\overline{\Delta }^{+} & f(\alpha )(p_0+\varepsilon _{\mathbf{p}%
})^{1-\alpha }
\end{array}
\right)
\end{equation}

\begin{equation}
\widehat{\widetilde{\Delta }}=\left( 
\begin{array}{cc}
0 & \widetilde{\Delta } \\ 
\widetilde{\Delta }^{+} & 0
\end{array}
\right)
\end{equation}
which can be written as:

\begin{equation}
S_{eff}\left( \widetilde{\Delta }^{+},\widetilde{\Delta }\right)
=-iTr\sum_{l=1}^\infty \frac 1l\left[ \widehat{G}\widehat{\widetilde{\Delta }%
}\right] ^l
\end{equation}
with:

\begin{equation}
\widehat{G}\left( p_0,\mathbf{p}\right) =\frac 1{f^2(\alpha )\left(
p_0^2-\varepsilon _{\mathbf{p}}^2\right) ^{1-\alpha }-\left| \overline{%
\Delta }\right| ^2}\left( 
\begin{array}{cc}
0 & -\widetilde{\Delta } \\ 
-\widetilde{\Delta }^{+} & 0
\end{array}
\right)
\end{equation}

We are interested in quadratic terms in $\widetilde{\Delta }$ and we will
take the approximation

\begin{equation}
S_{eff}\left( \widetilde{\Delta }^{+},\widetilde{\Delta }\right)
=S_{eff}^{(2)}\left( 0\right) +S_{eff}^{(2)}\left( \mathbf{q}\right)
\end{equation}
which contains the quadratic contributions. The first term in eq.( 2.102)
has the form 
\[
S_{eff}^{(2)}\left( 0\right) =\frac 12iTr\frac 1{f^2(\alpha )\left(
p_0^2-\varepsilon _{\mathbf{p}}^2\right) ^{1-\alpha }-\left| \overline{%
\Delta }\right| ^2}\left( \overline{\Delta }^2\widetilde{\Delta }^{+}%
\widetilde{\Delta }^{+}+\overline{\Delta }^{+2}\widetilde{\Delta }\widetilde{%
\Delta }+2\left| \overline{\Delta }\right| ^2\left| \widetilde{\Delta }%
\right| ^2\right) 
\]

\begin{equation}
+\frac 12iTr\frac 1{f^2(\alpha )\left( p_0^2-\varepsilon _{\mathbf{p}%
}^2\right) ^{1-\alpha }-\left| \overline{\Delta }\right| ^2}2\left| 
\widetilde{\Delta }\right| ^2
\end{equation}
which will be approximated, taking in the dominator $\overline{\Delta }%
\approx 0$ as:

\begin{equation}
S_{eff}^{(2)}\left( 0\right) \cong \frac 12iTr\frac 1{f^2(\alpha )\left(
p_0^2-\varepsilon _{\mathbf{p}}^2\right) ^{1-\alpha }}\left( \overline{%
\Delta }^2\widetilde{\Delta }^{+}\widetilde{\Delta }^{+}+\overline{\Delta }%
^{+2}\widetilde{\Delta }\widetilde{\Delta }+2\left| \overline{\Delta }%
\right| ^2\left| \widetilde{\Delta }\right| ^2\right)
\end{equation}
the last term giving no contribution to the renormalized coupling constant.
Following the same approximation we calculated $S_{eff}^{(2)}\left( \mathbf{q%
}\right) $ as:

\[
S_{eff}^{(2)}\left( \mathbf{q}\right) =\frac 12iTr\frac 1{f^2(\alpha
)(p_0-\varepsilon _{\mathbf{p}})^{1-\alpha }(p_0+q_0-\varepsilon _{\mathbf{%
p+q}})^{1-\alpha }}\widetilde{\Delta }\widetilde{\Delta }^{+} 
\]

\begin{equation}
+\frac 12iTr\frac 1{f^2(\alpha )(p_0-\varepsilon _{\mathbf{p}})^{1-\alpha
}(p_0+q_0-\varepsilon _{\mathbf{p+q}})^{1-\alpha }}\widetilde{\Delta }^{+}%
\widetilde{\Delta }
\end{equation}

>From eq. (2.104) and (2.105) we have

\[
L^{(2)}(0)=-\frac{B(1/2,3/2-2\alpha )}{4\pi f^2(\alpha )}(2m)^{3-4\alpha
}\times 
\]

\begin{equation}
\int \frac{d^2\mathbf{p}}{(2\pi )^2}\frac 1{(p^2+m\varepsilon _a)^{3-4\alpha
}}\left( \overline{\Delta }^2\widetilde{\Delta }^{+}\widetilde{\Delta }^{+}+%
\overline{\Delta }^{+2}\widetilde{\Delta }\widetilde{\Delta }+2\left| 
\overline{\Delta }\right| ^2\left| \widetilde{\Delta }\right| ^2\right)
\end{equation}
and

\[
L^{(2)}\left( \mathbf{q}\right) =-\frac{\sin (\pi (1-\alpha ))B(\alpha
,\alpha )}{4\pi f^2(\alpha )}m^{1-2\alpha }\int \frac{d^2\mathbf{p}}{(2\pi
)^2}\frac 1{\left( p^2+m\varepsilon _a+q_0m+q^2/4\right) ^{1-2\alpha }} 
\]

\begin{equation}
-\frac{\sin (\pi (1-\alpha ))B(\alpha ,\alpha )}{4\pi f^2(\alpha )}%
m^{1-2\alpha }\int \frac{d^2\mathbf{p}}{(2\pi )^2}\frac 1{\left(
p^2+m\varepsilon _a-q_0m+q^2/4\right) ^{1-2\alpha }}
\end{equation}

The integrals from eq. (2.106) and (2.107) can be performed using the formula

\begin{equation}
\int_{\mathbf{p}}\frac 1{(p^2+A^2)^N}=\frac{\Gamma (N-d/2)}{(4\pi
)^{d/2}\Gamma (N)}\frac 1{\left( A^2\right) ^{N-d/2}}
\end{equation}
and we obtain

\[
L^{(2)}=-\frac m{16\pi ^2f^2(\alpha )}\frac{2^{2-4\alpha }}{1-2\alpha }\frac{%
B(1/2,3/2-2\alpha )}{\varepsilon _a^{2-4\alpha }}\left( \overline{\Delta }^2%
\widetilde{\Delta }^{+}\widetilde{\Delta }^{+}+\overline{\Delta }^{+2}%
\widetilde{\Delta }\widetilde{\Delta }+2\left| \overline{\Delta }\right|
^2\left| \widetilde{\Delta }\right| ^2\right) 
\]

\[
+\frac m{16\pi ^2f^2(\alpha )}\frac{\sin (\pi (\alpha -1))}{2\alpha }\frac{%
B(\alpha ,\alpha )}{\left( \varepsilon _a+q_0+q^2/4m\right) ^{-2\alpha }}%
\widetilde{\Delta }\widetilde{\Delta }^{+} 
\]

\begin{equation}
+\frac m{16\pi ^2f^2(\alpha )}\frac{\sin (\pi (\alpha -1))}{2\alpha }\frac{%
B(\alpha ,\alpha )}{\left( \varepsilon _a-q_0+q^2/4m\right) ^{-2\alpha }}%
\widetilde{\Delta }^{+}\widetilde{\Delta }
\end{equation}

Using the approximation:

\begin{equation}
\left( \varepsilon _a-q_0\pm q^2/4m\right) ^{2\alpha }=\varepsilon
_a^{2\alpha }+2\alpha \varepsilon _a^{2\alpha -1}\left( \pm q_0+\frac{q^2}{4m%
}\right)
\end{equation}
and using the notation

\begin{equation}
\widetilde{\Psi }=\left( 
\begin{array}{c}
\widetilde{\Delta } \\ 
\widetilde{\Delta }^{+}
\end{array}
\right)
\end{equation}
we obtain from eq. (2.109)

\begin{equation}
L^{(2)}=\frac m{16\pi ^2f^2(\alpha )}\sin (\pi (1-\alpha ))B(\alpha ,\alpha
)\varepsilon _a^{2\alpha -1}\frac 12\widetilde{\Psi }^{+}M\widetilde{\Psi }
\end{equation}
where

\begin{equation}
M=\left( 
\begin{array}{cc}
q_0-\frac{q^2}{2m_B}-\mu _0 & -\mu _0 \\ 
-\mu _0 & -q_0-\frac{q^2}{2m_B}-\mu _0
\end{array}
\right)
\end{equation}
$m_b=2m$ being the boson mass and $\mu _0$ the chemical potential

\begin{equation}
\mu _0=\frac 1{f^2(\alpha )}\frac{2^{2-4\alpha }B(1/2,3/2-2\alpha )}{%
(1-2\alpha )\sin (\pi (1-\alpha ))B(\alpha ,\alpha )}\left| \overline{\Delta 
}\right| ^2\varepsilon _a^{2\alpha -1}
\end{equation}
the velocity $c_0$ of the sound mode is:

\begin{equation}
c_0=\frac{\mu _0}{m_b}=\frac 1{f^2(\alpha )}\frac{2^{2-4\alpha
}B(1/2,3/2-2\alpha )}{(1-2\alpha )\sin (\pi (1-\alpha ))B(\alpha ,\alpha )m}%
\left| \overline{\Delta }\right| ^2\varepsilon _a^{2\alpha -1}
\end{equation}
and the repulsive interaction $\lambda _{0b}$ between pairs is:

\begin{equation}
\lambda _{0b}=\frac{\pi ^2}m\frac{2^{4-4\alpha }B(1/2,3/2-2\alpha )}{%
(1-2\alpha )[\sin (\pi (1-\alpha ))B(\alpha ,\alpha )]^2}
\end{equation}

We mention that $\lim_{\alpha \rightarrow 0}\lambda _{0b}(\alpha )=2\pi /m$
a result identical to the result obtained in Ref. [60] for the 2D case.

\vspace{0.1in}

\section{Summary of the Results}

\vspace{0.1in}

$\bullet $ In the first part of this chapter we performed a simple
calculation of the pseudogap due to the interaction between electrons and
magnetic fluctuations. We showed that for the 2D systems, in one loop
approximation the self-energy of the electronic excitation can be expressed
analytically by the structure factor $S(\mathbf{q},T)$. The expression for
the real and imaginary part of the self-energy given by eqs. (2.17) and
(2.18) are very usefully because their behavior as function of $\omega $ and 
$\mathbf{q}$ can be easily controlled and predicted if the form of $S(%
\mathbf{q},T)$ is known. This calculation make transparent the analytical
approach first proposed in Ref.[22-25] using the self consistent treatment.

We considered in the susceptibility $\chi (\mathbf{q},\omega )$ the
diffusive contribution of the spin waves and we showed that if the condition
expressed in eq. (2.21) is satisfied the spin-wave does not change the
occurrence of the pseudogap predicted by the coupling with low energy spin
fluctuations. The energy scale has been obtained from the condition $%
G(\omega )\sim 1/\omega $ ( $\Sigma (\omega )\sim 1/\omega $ ) and in fact
this condition gives for the coherence length of the magnetic fluctuations
an expression of the form $\xi (T)\sim \exp [C/T]$ used in [25] in the
renormalized classical regime, and which is appropriate also for a 2D
magnetic system [36]. Recently the RNG equations have been applied for the
proximity of the Lifshitz point [37] .

$\bullet $ We also tried to consider the effect of the anisotropy in the
magnetic fluctuations and following a calculation similar to that of Ref.
[34] we calculated a pseudogap which is temperature dependent. This
dependence is even more complicated because in different regimes the
coherence length $\xi (T)$ for the magnetic fluctuations is different. Using
different $T$ dependence for $\xi (T)$ we showed that even for $\xi (T)\sim
\exp [C/T]$ there is a temperature dependence in the pseudogap which is not
certain from experimental point of view [38].

We mention that the key problem of such model remain the matching between
the pseudogap, which in this case has a magnetic origin, and the
superconducting gap which should have a d-wave pairing origin. In order to
have a picture of this point we propose that the following qualitative
behavior of the phase diagram. The total gap $\Delta (T)$ is defined as:

\begin{equation}
\Delta (T)=\left\{ 
\begin{tabular}{ll}
$\Delta _s(T)+$ $\Delta _{pg}(T)$ & , $T<T_c$ \\ 
$\Delta _{pg}(T)$ & , $T_c<T<T^{*}$%
\end{tabular}
\right.
\end{equation}
where $T_c$ is the superconducting critical temperature and $T^{*}$ is the
pseudogap appiaring temperature. According to this picture the measured gap
below the superconducting critical temperature has two contributions, one
from the superconducting gap and the other one from the fluctuation gap.

$\bullet $ In the second part of this chapter we showed that a temperature
dependent pseudogap appears in a non-Fermi liquid superconductor due to the
interaction between electrons and the fluctuations of the order parameter
amplitude. The mode-mode coupling, valid in the weak coupling approximation
can give relevant results,even for the intermediate coupling studied by
Levin group [39]. The model recently applied by Norman et.al. [40] can be
applied for the spin fluctuation model proposed by Chubukov [9] in order to
study the temperature dependence of the pseudogap. In Ref. [40] the filling
in of the pseudogap due to the increment of the temperature is given by the
broadening in the self-energy and is proportional to $T-T_c$. A similar
broadening effect proportional to $\widetilde{\tau }(\alpha )$ was obtained
in our model and this can be seen very easily from eq. (2.77) if in the
electronic Green function we take the limit $q=0$.

Recently such a model for the Fermi liquid superconductor has been studied
by Kristoffel and Ord [41] and their temperature dependence is different
than our result. However, we mention that according to their model these
authors have to obtain a result similar to the results given in Ref.[25].
The difference is given by the method of performing the integral over $%
\mathbf{q}$ which is not correct in Ref. [41].

Recently Preosti et.al. [34] generalized the method given in Ref. [25]
taking into consideration the anisotropy in the dynamic susceptibility due
to the interplane pairing. From the temperature dependence of the pseudogap
shown in Fig. 2.3 one can see that there is a narrow domain of temperature
where this dependence is in fact in agreement with the recent data from Ref.
[41]. Increasing the temperature will lead the system to the pure classical
regime, where the pseudogap is constant, result in agreement with Ref. [25].

The model analyzed in Ref. [25] and [41] use a constant coupling between
electrons and overdamped fluctuations describe by the t-matrix or a $\chi (%
\mathbf{q},\omega )$ containing an imaginary part. Recently Tchernyshyov
[42] showed that using a better approximation for the t-matrix the decay of
the Cooper pairs is negligible and a bosonic propagator without damping of
excitations gives a pseudogap in the normal state. This idea is also
interesting for our model. This simple model neglected the vertex
corrections in the electronic self-energy, which for the non-Fermi liquid
superconductor are very singular and has to be considered in the transport
theory.

At the present time it is generally accepted that the pseudogap appears only
in the underdoped cuprates. This fact is reflected in our scenario by the
dependence of $\Delta _{pg}$ of $\alpha $ which appears as a parameter of
the model. However it was showed [44] that the correlation length depends on 
$\alpha $ and we can attend the overdamped regime by variation of $\alpha $.
This demonstrate that such a model is appropriate for the description of the
pseudogap in a non-Fermi-liquid model.

$\bullet $In the last part of this chapter using the field-theoretical
methods we studied the crossover between BCS and BEC in a non-Fermi liquid.
The weak coupling case lead to the same results as in the mean field like
models. In the strong coupling limit we showed that the pairs form a Bose
gas with a repulsive coupling constant which is controlled by $\alpha $

\chapter{Renormalization Group Approach of Itinerant Electron System
near the Lifshitz Point}

\vspace{0.1in}

\section{Introduction}

\vspace{0.1in}

The occurrence of the non-Fermi behavior in the systems of fermions coupled
to a critical fluctuations mode has been suggested in connection with the
neutron experiments [1-2] and studied in the framework of many body theory
[3-4] in the case of two-dimensional (2D) and three dimensional (3D) models.
Recent experiments on the heavy fermions systems showed also a non-Fermi
behavior of these materials at low temperatures and it was associated with
the proximity of quantum critical point (QCP). The most studied example [5]
is $CeCu_{6-x}Au_x$ where at the QCP $x=0.1$ the resistivity increases
linearly with temperature T over a wide range of T and the specific heat $%
C(T)$ is proportional to $T\ln (T_0/T)$. This behavior has been explained
[6] by the coupling of 3D fermionic excitations to the 2D critical
ferromagnetic fluctuations near the QCP. The inelastic neutron scattering
measurements performed on this materials [7-8] showed the following new
points in the behavior of this material

$\bullet $The inelastic neutron scattering data can be fitted with a
susceptibility of the form: $\chi ^{-1}=C^{-1}[f(q)+(aT-i\omega )^\alpha ]$
where $\alpha =4/5$ and not $1$ as is predicted by the mean field
approximation

$\bullet $The quadratic stiffness vanishes, fact which shows that we are
dealing with a quantum Lifshitz point (QLP)

$\bullet $The peaks for $x=0.2$ and $x=0.3$ can be considered as 2D
precursor of 3D order

$\bullet $The scaling analysis showed [7] that $\gamma (T)=C(T)/T$ has the
form

\begin{equation}
\gamma (T)\sim T^{(D-1/2)a/2-1}
\end{equation}

which for $D=3$ and $\alpha =4/5$ gives a temperature independent value.
This analysis has been performed taking as the most important contribution
to $\chi $ the form containing $\omega ^\alpha $ and the q-dependence of the
form $f(q)=Dq_{\parallel }^2+Cq_{\perp }^4$ where $\alpha =2/z$, $z$ being
the critical exponent from the dynamical critical phenomena [9]. In this
chapter we will show that using the Hertz [10] renormalization group method
(RNG) extended for $T\neq 0$ by Millis [11] we can obtain the $\ln (T_0/T)$
term as a quantum correction to the classical results expressed by eq. (3.1).

In the next section the renormalization group studies of the Gausian fixed
point of magnetic transitions in the metallic phase will be use. This
treatment of the Gaussian fixed point essentially lead to equivalent results
to the self-consistent renormalization approximation [12]. For the case of
ferromagnetic system with isotropic Fermi surface we have for the free
susceptibility the following expansion:

\begin{equation}
\chi (\mathbf{k},i\omega _n)=N(0)\left[ 1-\frac 13\left( \frac{\mathbf{k}}{%
2k_F}\right) ^2-\frac \pi 2\frac{\left| \omega _n\right| }{kv_F}+...\right]
\end{equation}

It should be noted that this expansion is possible only for $d>3$. In $d=1$
and $d=2$ the susceptibility is singular at $\left| \mathbf{k}\right| =2k_F$%
, even for an isotropic Fermi surface. Recently the nonanalyticy of $\chi (%
\mathbf{k},i\omega _n)$ in $\mathbf{k\ }$for $d<3$ in the leading order was
considered more seriously [14] which led to the conclusion that for the
ferromagnetic case, the mean field description is incorrect at $d=2$ and $%
d=3 $ in contrast with the result from eq. (3.1).

In the antiferromagnetic case, $\chi (\mathbf{Q+k},i\omega _n)$ is similarly
expended for small $\mathbf{k}$ and $\omega _n$ if the nesting condition is
not satisfied and the spatial dimension satisfies $d\geq 3$. When the
nesting condition is satisfied, as in the case of Mott insulator state, this
type of simple expansion is not possible.

\vspace{0.1in}

\section{The Scaling Equations}

\vspace{0.1in}

The main idea of our model is contained in a modified form of the
susceptibility given in Ref. [7] as:

\begin{equation}
\chi ^{-1}\left( \mathbf{q},i\omega _n\right) =C\left( f(q)+\delta ^\alpha
+(aT)^\alpha +\left| \omega \right| ^\alpha \right)
\end{equation}
where $f(q)$ is a smooth function of the wave vector, $\delta $ is the
control parameter (pressure, doping) measuring the distance from QCP and $%
\alpha $ is the phenomenological exponent. The real part of this expression
satisfies for $\omega =0$ the relation:

\begin{equation}
\chi ^{^{\prime }-1}\left( \mathbf{q},T\right) -\chi ^{^{\prime }-1}\left(
0,T\right) =C(aT)^{-\alpha }
\end{equation}
which is identical to the relation satisfied by the form proposed in Ref.
[7]. The imaginary part of the susceptibility expressed by eq. (3.3) has the
form

\begin{equation}
\chi ^{\prime \prime }\left( \omega ,T\right) =T^{-\alpha }g\left( \frac
ET,\frac \delta T\right)
\end{equation}
where:

\begin{equation}
g(y,x)=C\frac{\sin \left[ \arctan \left( y^\alpha \frac{\sin (\pi \alpha /2)%
}{1+x^\alpha +y^\alpha \cos (\pi \alpha /2)}\right) \right] }{\left[ \left(
1+x^\alpha +y^\alpha \cos (\pi \alpha /2)\right) ^2+y^{2\alpha }\sin (\pi
\alpha /2)\right] ^{1/2}}
\end{equation}
and $f(q)=0$. We mention that for $\delta =0$ and $\alpha =1$ we reobtain
the mean field results and it can be showed numerically that the scaling
function $g(y,x)$ has the same form with the scaling function $g(y)$
obtained in Ref. [7].

Using these considerations we consider that in the low temperature
approximation the interacting Fermi system can be describe by the effective
action:

\begin{equation}
S_{eff}\left( \Phi \right) =S_{eff}^{(2)}(\Phi )+S_{eff}^{(4)}(\Phi )
\end{equation}
where:

\begin{equation}
S_{eff}^{(2)}(\Phi )=VT\sum_n\int \frac{d^3\mathbf{q}}{(2\pi )^3}\left[
\delta ^\alpha +\left| \omega _n\right| ^\alpha +q_{\parallel }^2+Dq_{\perp
}^2+q_{\perp }^4\right] \left| \Phi (q,\omega _n)\right| ^2
\end{equation}

\begin{equation}
S_{eff}^{(4)}(\Phi )=uVT^3\sum_{n_i}\prod_{i=1}^4\frac{d^3\mathbf{q}_i}{%
(2\pi )^3}\Phi (q_i,\omega _n)\delta \left( \sum_{i=1}^4\mathbf{q}_i\right)
\delta \left( \sum_{i=1}^4\omega _{n_i}\right)
\end{equation}
with $\omega _n$ a bosonic frequency, $u>0$ is the coupling constant and $D$
is the stiffness of the fluctuations.

In order to calculate specific heat we will use the Gaussian form of the
free energy obtained from eq. (3.8) as:

\begin{equation}
F=\int_0^1\frac{dz}{2\pi }\int_0^1\frac{d^2\mathbf{q}_{\parallel }}{(2\pi )^2%
}\int_0^1\frac{dq_{\perp }}{2\pi }\coth \frac z{2T}\arctan \frac{A\sin
\theta }{A\cos \theta +q_{\parallel }^2+Dq_{\perp }^2+q_{\perp }^4}
\end{equation}
where $\theta =\arctan (z/\delta ^\alpha )$ and $A^{-\alpha }=(\delta
^{2\alpha }+z^2)^{1/2}$. Using the transformations

\begin{equation}
\begin{array}{ccc}
\omega ^{\prime }=b^{2/\alpha }\omega & q_{\parallel }^{\prime
}=bq_{\parallel } & q_{\perp }^{\prime }=b^{1/2}q_{\perp } \\ 
T^{\prime }=b^{2/\alpha }T & \delta ^{\prime }=b^{2/\alpha }\delta & 
D^{\prime }=bD
\end{array}
\end{equation}
and following the same procedure as in Ref. [11] we obtain the scaling
equations:

\begin{equation}
\frac{dT(b)}{d\ln b}=\frac 2\alpha T(b)
\end{equation}

\begin{equation}
\frac{du(b)}{d\ln b}=\left( \frac 32-\frac 2\alpha \right) u(b)-u^2(n+8)f_2
\end{equation}

\begin{equation}
\frac{d\delta ^\alpha (b)}{d\ln b}=2\delta ^\alpha (b)+2u(b)(n+2)f_1
\end{equation}

\begin{equation}
\frac{dD(b)}{d\ln b}=D(b)
\end{equation}

\begin{equation}
\frac{dF(b)}{d\ln b}=\left( \frac 52+\frac 2\alpha \right) F(b)+f_3
\end{equation}
where $n$ is the number of the field component, $f_1=f_1[T(b),\delta ^\alpha
(b),D(b)]$, $f_2=f_2[T(b),\delta ^\alpha (b),D(b)]$ and $f_3=f_3[T(b)]$ are
complicated functions but can be approximated in the limit of weak
dependence of $\delta ^\alpha (b)$ and $D(b)$ for $\delta ^\alpha
(b),D(b)\ll 1$. If we are near a QCP which is usually at very low
temperature the scaling equations will be linearized, keeping only the
linear term in the coupling constant. The renormalization procedure is
stopped at

\begin{equation}
\delta ^\alpha (b)=1
\end{equation}

>From the linearized eqs. (3.12)-(3.14) we get:

\begin{equation}
T(b)=Tb^{2/\alpha }
\end{equation}

\begin{equation}
u(b)=ub^{3/2-2/\alpha }
\end{equation}

\begin{equation}
\delta ^\alpha (b)=e^{2\ln b}\left[ \delta ^\alpha +2u(n+2)\int_0^{\ln
b}dxe^{-x(1/2+2/\alpha )}f_1\left( Te^{2x/\alpha }\right) \right]
\end{equation}

These equations will be analyzed in two cases which are in fact the low
temperature and high temperature regimes.

In low temperature limit the integral from eq. (3.20) can be approximated as:

\begin{equation}
I=\int_0^{\ln b}dxe^{-x(1/2+2/\alpha )}f_1(Te^{2x/\alpha })\cong \frac{f_1(0)%
}{1/2+2/\alpha }
\end{equation}
and eq. (3.20) becomes

\begin{equation}
\delta ^\alpha (b)=b^2\left[ \delta ^\alpha +\frac{2u(n+2)f_1(0)}{%
1/2+2/\alpha }+2Bu(n+2)T^{1+\alpha /4}\right]
\end{equation}

If we introduce the parameter

\begin{equation}
r_\alpha =\delta ^\alpha +\frac{2u(n+2)f_1(0)}{1/2+2/\alpha }
\end{equation}
the low temperature regime is defined by

\begin{equation}
T(b)\ll 1
\end{equation}
and the high temperature regime is defined by

\begin{equation}
T(b)\gg 1
\end{equation}

>From eq. (3.22) using the condition $\delta ^\alpha (\overline{b})=1$ we get
the condition for the low temperature regime as

\begin{equation}
\frac T{r_\alpha ^{1/\alpha }}\ll 1
\end{equation}

In the high temperature regime defined now by the inequation (3.26)
reversed, we approximate the function $f_1(T)$ as $f_1(T)\cong CT$ and
introduce the new variable $v=uT$. The linearized scaling equations are:

\begin{equation}
\frac{dv(b)}{d\ln b}=\frac 32v(b)
\end{equation}

\begin{equation}
\frac{d\delta ^\alpha (b)}{d\ln b}\cong 2\delta ^\alpha (b)+2Cv(b)(n+2)
\end{equation}
and the gaussian behavior appears if $\delta ^\alpha (b)=1$ and $v(b)\ll 1$.
The general solution of eqs. (3.27) and (3.28) have the form

\begin{equation}
v(b)=\overline{v}e^{3\ln b/2}
\end{equation}

\begin{equation}
\delta ^\alpha (b)=e^{2\ln b}\left[ \overline{\delta }^\alpha +2\widetilde{C}%
\overline{v}\right] -4\widetilde{C}\overline{v}e^{3\ln b/2}
\end{equation}
with $\widetilde{C}=2(n+2)C$.

The initial conditions have been determined following the procedure from
Ref. [11]

\begin{equation}
\overline{\delta }^\alpha =T^{-\alpha }\left[ r_\alpha +2Bu(n+2)T^{1+\alpha
/4}\right]
\end{equation}
where $B$ is a constant defined in Ref.[11]. The condition $v(b)\ll 1$ is
defined if:

\begin{equation}
R=\frac{uT}{\left[ r_\alpha +2(B+C)u(n+2)T^{1+\alpha /4}\right] ^{3/4}}\ll 1
\end{equation}
which is in fact the condition for the validity of the Gaussian model
(Ginsburg criterion). The critical temperature is well approximated by:

\begin{equation}
T_c=\left[ \frac{r_\alpha }{2(B+C)u(n+2)}\right] ^{4/(4+\alpha )}
\end{equation}

This result for $T_c(\delta )$ can be obtained from eq. (3.33) as:

\begin{equation}
T_c\sim \delta ^{4\alpha /(4+\alpha )}
\end{equation}
and for $\alpha =4/5,T_c\sim \delta ^{0.67}$.

\vspace{0.1in}

\section{The Specific Heat}

\vspace{0.1in}

In order to calculate the specific heat we will use eq. (3.10) for the free
energy F. Neglecting in the lower approximation the second term we obtain

\begin{equation}
F(T)=F(b)b^{-2/\alpha -5/2}
\end{equation}

The exact solution of eq. (3.16) has the form

\begin{equation}
F(b)=b^{2/\alpha +5/2}\int_0^{\ln b}dxe^{-(2/\alpha +5/2)x}f_3(Te^{2x/\alpha
})
\end{equation}

In order to calculate the temperature dependence of the free energy
expressed by eq. (3.36) we consider the variable $x$ in the domains

\begin{equation}
0<x<\frac \alpha 2\ln \frac 1T
\end{equation}

\begin{equation}
\frac \alpha 2\ln \frac 1T<x<\ln b^{*}
\end{equation}
where $b^{*}$ is defined by $\delta ^\alpha (b^{*})=1$ and from eq. (3.17)
was calculated as $b^{*}=T^{-\alpha /2}$. Following Ref.[11] in the first
domain $f_3(T)\cong C_3T$ and in the second domain $f_3(T)\cong DT$. Using
these approximations we obtain from eq. (3.36)

\begin{equation}
b^{-2/\alpha -5/2}F(b)=\frac \alpha 2T^{1+5\alpha /4}\left[
C_3\int_T^1dT_1T_1^{-5\alpha /4}+D\int_1^{Tb^{*2/\alpha }}dT_1T_1^{-5\alpha
/4}\right]
\end{equation}
where $T_1=T\exp [2x/\alpha ]$. If we take $\alpha =4/5$ from eq. (3.39) we
calculate

\begin{equation}
F(T)=\frac 25C_3T^2\ln \frac 1T+\frac 25DT^2-\frac 25DTb^{*-5/2}
\end{equation}
and $\gamma (T)=C_v/T$ as:

\begin{equation}
\gamma (T)=\gamma _{c0}+\overline{\gamma }\ln \frac 1T+O\left( \frac
1{T^2}\right)
\end{equation}
a result which shows that using RNG for the phenomenological model with $%
\alpha =4/5$ we obtain the $\ln (1/T)$ term in $\gamma (T)$which is in fact
done by the non-Fermi behavior of the model.

Recently Ramazashvilli [13] used the same method studying QLP for such a
model with $\alpha =1$. Our results are consistent with the results from
Ref. [13], excepting the specific heat coefficient which has $T^{1/4}$
dependence. The phase diagram can be calculated from the free energy

\begin{equation}
F\sim \left( \delta ^\alpha +Dq^2+q^4\right) \Phi ^2
\end{equation}
and for $D>0$, $\delta ^\alpha $ we get an ordered phase at $q=0$. A second
ordered phase can be obtained for $D<0,$ $q=\pm (D/2)^2$ separates the two
ordered phases from disordered phase. The importance of the $q_{\parallel
}^2 $ and $q_{\perp }^4$ terms in the susceptibility can be also discussed
as in Ref. [13] and we get the same result, the only difference being that $%
\delta (b)$ has to be replace with $\delta ^\alpha (b)$.

\vspace{0.1in}

\section{Summary of the Results}

\vspace{0.1in}

$\bullet $We calculated the specific heat dependence of a 3D itinerant
electron system near a Lifshitz point using the phenomenological model
considering the susceptibility of the form $\chi ^{-1}\sim f(q)+\delta
^\alpha +\left| \omega \right| ^\alpha +T^\alpha $. This form describes a
similar behavior with the phenomenological model proposed in Ref. [7,8] and
using the $T\neq 0$ RNG proposed by Millis we showed that the specific heat
presents a $\ln T$ contribution for $\alpha =4/5$ which is typical behavior
for a non-Fermi system.

Our results can be considered as generalization of the recent results
obtained in Ref. [13] where the exponent $\alpha $ has the value $\alpha =1$%
. The first calculation [11] of the specific heat using RNG methods near a
QCP with zero critical temperature used for the dynamic exponent $z=2,z=3$
or $z=4$ but no $\ln T$ dependence has been obtained.

\chapter{Conclusion}

\vspace{0.1in}

We conclude this thesis presenting the original results obtained and
mentioning the unsolved problems in this field.

\textit{Chapter 1:}

(1) The temperature dependence of a marginal Fermi liquid has been
calculated . We showed that the expected $T\ln T$ correction is
characteristic for the low temperature domain. The high temperature domain
has a supplementary correction. The results are in agreement with the
non-Fermi behavior of some metallic systems in the low temperature
domain.(results contained in: \textit{M.Crisan, C.P. Moca Journal of
Superconductivity }\textbf{9} 49 (1996))

(2) We calculated the self-energy at $T=0$ for a two dimensional fermionic
system with hyperbolic dispersion. The existence of the saddle points in the
energy gives rise to a marginal behavior, a result which has been obtained
by numerical calculations. We present the many-body calculation of the
self-energy. We showed that even in a (RPA) approximation this model present
a marginal Fermi liquid behavior.(results contained in: \textit{C.P. Moca,
M. Crisan Journal of superconductivity} \textbf{10} 3 (1997) )

(3) In order to give a stronger support to this model we adopted the
Renormalization Group Method for this model.. The calculation of the wave
function renormalization constant $Z$ will be performed using the ''poor
man's renormalization'' method and we will show that $Z\rightarrow 0$ with
the infrared cut-off $\Lambda $ as $Z(\Lambda )=\Lambda ^{\zeta }$ where $%
\zeta $ is a constant. (results contained in: \textit{M. Crisan, C. P. Moca
Modern Physics Letters} \textbf{B9} 1753 (1995))

The last experimental results and the theoretical investigation showed that
the non-Fermi behavior is associated with the two novel features of these
systems, namely the occurrence of the \textbf{pseudogap} and the proximity
of a \textbf{Quantum Phase Transition. }This problems have been studied by
different authors for a normal Fermi-liquid. However we adopted the Anderson
non-Fermi model because it can be reduced to the normal model taking the
parameter $\alpha $ equal zero.

\textit{Chapter 2}

(4) The electronic self-energy due to the electron-spin interaction is
calculated using the one-loop approximation for the two dimensional system
and quasi-two dimensional (anisotropic) model. We analyzed the relevance of
the diffusive modes and the temperature dependence of the magnetic
correlation length for a possible temperature dependence of the pseudogap.
The results contained in this section are in the spirit of the method
presented by the Tremblay group for the normal Fermi systems. (results
contained in: \textit{C.P. Moca, I.Tifrea, M. Crisan (accepted for
publication in Journal of Superconductivity)} )

(5) We study the influence of the amplitude fluctuations of a non-Fermi
superconductor on the energy spectrum of the two-dimensional Anderson
non-Fermi system. In order to perform such a calculation we had to elaborate
Schmid self-consistent model of fluctuations in BCS superconductor for the
fluctuations in an Anderson non-Fermi superconductors. The new propagator of
fluctuations have been calculated and in the limit of $\alpha =0$ it gives
the results from BCS. These fluctuations are also classical and give a
temperature dependence in the pseudogap induced in the fermionic excitations
of the Anderson model. (results contained in: \textit{M. Crisan, C.P. Moca,
I. Tifrea Phys.Rev.} \textbf{B59} 14680 (1999))

In order to study the possibility of the superconducting state in the
Anderson model we also studied the problem formulated by the Nozieres and
Schmitt-Rink who considered the crossover between weak (BCS) and strong
coupling (BEC) using a Fermi liquid model.

(6) Using the field-theoretical methods we studied the evolution from BCS
description of a non-Fermi superconductor to that of Bose-Einstein
condensation (BEC) in one loop approximation. We showed that the repulsive
interaction between composite bosons is determined by the exponent $\alpha $
of the Anderson propagator in a two dimensional model. For $\alpha \neq 0$
the crossover is also continuous and for $\alpha =0$ we obtain the case of
the Fermi liquid.(results contained in:I. Tifrea, \textit{C.P. Moca, M.
Crisan (presented at the 6}$^{th}$ \textit{International Conference,
Materials and Mechanisms of Superconductivity and High Temperature
Superconductors, February 20-25, 2000, Houston, Texas, USA , to be published
in Physica C})

The proximity of a Quantum Phase Transition, recently verified by many
experimental results, has been studied using the RNG method for the
phenomenological model proposed by the group of Hilbert von Lohneysen.

\textit{Chapter 3}

(7) Using the renormalization group approach proposed by Millis for the
itinerant electron systems, in the case of $d=z=2.5$ we calculated the
specific heat coefficient $\gamma (T)$ for the magnetic fluctuations with
susceptibility $\chi ^{-1}\sim \delta ^\alpha +\left| \omega \right| ^\alpha
+f(q)$ near a Lifshitz point. The constant value for $\alpha =4/5$ and the
logarithmic temperature dependence, specific heat for the non-Fermi
behavior, have been obtained in agreement with the experimental
data.(results contained in: \textit{C.P. Moca, I. Tifrea, M. Crisan
(accepted for publication in Phys. Rev. \textbf{B})} )

\textbf{Open problems}:

In spite of fact that the non-Fermi model proposed by Varma et.al. and
Anderson may explain many experimental results the new very accurate
measurements showed that there are still open problems which have to be
explained even in a semi-phenomenological model. In this respect a non-Fermi
liquid model have to explain quantitatively the

$\bullet $ peak-dip-hump structure of the ARPES lineshape and tunnel
spectroscopy.

$\bullet $ the pseudogap spectrum in vortex core of underdoped cuprates
superconductors

$\bullet $ the existence of resonances in neutron scattering

$\bullet $ marginal or non-Fermi liquid like relaxation rates in optical
conductivity

$\bullet $ nature of the magnetic fluctuations and the magnetic properties
of the non-Fermi liquid as well as the transport in (pseudo) two-dimensional
non-Fermi systems. In this problem we got some new results studing the
electron-hole channel for the non-Fermi Anderson Model (results contained
in: \textit{M. Crisan, C.P. Moca, I. Tifrea (to be published )} )

\vspace{0.5cm}


\begin{thebibliography}{99}
\bibitem{}  C.M.Varma, P.B. Littlewood, S. Schmitt-Rink, E. Abrahams, A.E.
Ruckenstein \textit{Phys.Rev.Lett.} \textbf{63} 1996 (1989)

\bibitem{}  G. Kotliar, E. Abrahams, A.E. Ruckenstein , C.M. Varma, P.B.
Littlewood, S. Schmitt-Rink \textit{Europhys.Lett.} \textbf{15} 655 (1991)

\bibitem{}  P.B. Littlewood, J. Zaaneu, G. Apple, H. Monien \textit{%
Phys.Rev. }\textbf{B48} 487 (1993)

\bibitem{}  M.L. Horbach, F.L.J. Vos and W.van Sarlos \textit{Phys.Rev.} 
\textbf{B48} 4061 (1993)

\bibitem{}  C. Bendle, P.Hertel, J. Apple \textit{Phys.Rev.} \textbf{B45}
8062 (1992)

\bibitem{}  M.Y. Reizer \textit{Phys.Rev.} \textbf{B40} 11571 (1984)

\bibitem{}  P.W. Anderson \textit{Physica} \textbf{2} 1 (1965)

\bibitem{}  R. Balian, D.R. Fredkin \textit{Phys.Rev.Lett.}\textbf{\ 15} 480
(1965)

\bibitem{}  N.F. Berk, J.R. Schriffer \textit{Phys.Rev.Lett.} \textbf{17}
433 (1966)

\bibitem{}  D.J. Amit, J.W. Kane, H. Wegner \textit{Phys.Rev.} \textbf{175}
313 (1968)

\bibitem{}  G.M. Carneiro, C.P. Petrick\ \textit{Phys.Rev.} \textbf{B37} 442
(1988)

\bibitem{}  R. Chandhury \textit{Can.J.Phys.} \textbf{73} 497 (1995)

\bibitem{}  F. Steglich in Material and Mechanisms of Superconductivity 
\textit{(North Holland Amsterdam)(1998)}

\bibitem{}  D.M.Newns, H.R.Krishamurthy, P.C. Patnaik, C.C. Tsuei, C.L.Kane 
\textit{Phys.Rev.Lett.} \textbf{69} 1264 (1992)

\bibitem{}  A. Millis, H. Monien, D. Pines Phys.Rev. B42 167 (1990) see also
B.P. Stjkovic, D. Pines \textit{Phys.Rev.Lett.}\textbf{\ 76} 811 (1996)

\bibitem{}  I. Grosu, M. Crisan \textit{Phys.Rev.}\textbf{\ B49} 1296 (1994)

\bibitem{}  H.H. Amin, P.C.E. Stamp \textit{Phys.Rev.Lett.} \textbf{77} 3017
(1996)

\bibitem{}  P. Bernard, L. Chen, A-M.S. Tramblay \textit{Phys.Rev.} \textbf{%
B47} 15127 (1989)

\bibitem{}  M. Crisan, L. Tataru \textit{Phys.Rev.} \textbf{B54} 1 (1996)

\bibitem{}  J. Friedel \textit{J. Phys. (Paris)} \textbf{48} 1787 (1987)

\bibitem{}  J. Labbe, J. Bok, \textit{EuroPhys.Lett.}\textbf{\ 2} 1225 (1987)

\bibitem{}  D.M.Newns, P.C. Pattnaik, C.C. Tsuei \textit{Phys.Rev.}\textbf{\
B43 }3075 (1991)

\bibitem{}  P.C. Pattnaik, C.L. Kane, D.M. Newns,C.C. Tsuei \textit{Phys.Rev.%
}\textbf{\ B45} 5714 (1992)

\bibitem{}  S. Gopalan, O. Gunnarson, O.K. Andersen \textit{Phys.Rev.}%
\textbf{\ B46} 11798 (1992)

\bibitem{}  J. Gonzales, F. Guineea \textit{EuroPhys.Lett.} \textbf{34} 711
(1996)

\bibitem{}  D.Coffey, C.P. Petrick \textit{Phys.Rev.} \textbf{B37} 442 (1988)

\bibitem{}  D. Coffey, K.S. Bedell \textit{preprint 590-1993 Los Alamos
Centre for Materials Science}

\bibitem{}  B. Andraka, A.M. Tsvelik \textit{Phys.Rev.Lett.} \textbf{67}
2886 (1991)

\bibitem{}  S.Chakravarty, R.E. Norton, O.F. Syljnasen \textit{Phys.Rev.Lett.%
} \textbf{74} 1423 (1995)

\bibitem{}  P.M. Bares, X.G. Wen \textit{Phys.Rev.} \textbf{B48} 8636 (1993)

%\newpage\ 
\end{thebibliography}

\begin{thebibliography}{}
\bibitem{}  P.W. Anderson Science \textbf{235} 1196 (1987) ; \textit{%
Phys.Rev.Lett.} \textbf{64} 1839 (1990) ; \textbf{65} 2306 (1990)

\bibitem{}  S. Chakravarty, P.W. Anderson \textit{Phys.Rev.Lett.} \textbf{72}
3859 (1994)

\bibitem{}  X.G. Wen \textit{Phys.Rev.} \textbf{B42} 6623 (1990)

\bibitem{}  A.V. Balatsky \textit{Philos.Mag.Lett.}\textbf{\ 68} 251 (1993)

\bibitem{}  A. Sudbo \textit{Phys.Rev.Lett. }\textbf{74} 2575 (1995)

\bibitem{}  A. Sudbo, J.M. Wheatley \textit{Phys.Rev.}\textbf{\ B52} 6200
(1995)

\bibitem{}  V.N. Muthukumar, D. Sa, M. Sardar \textit{Phys.Rev.}\textbf{\ B52%
} 9647 (1995)

\bibitem{}  I. Grosu, I. Tifrea, M. Crisan, S. Yoksan \textit{Phys.Rev.} 
\textbf{B56} 8298 (1997)

\bibitem{}  A.V. Chubukov, J. Schmalian \textit{cond-mat/9711041}

\bibitem{}  V.G. Geshkenbein, L.B. Ioffe, A.I. Larkin \textit{Phys.Rev.}%
\textbf{\ B55} 3173 (1997)

\bibitem{}  S.C. Zang \textit{Science} \textbf{275} 1089 (1997)

\bibitem{}  P.A. Lee, X.G. Wen \textit{Phys.Rev.Lett.}\textbf{\ 78} 4111
(1997)

\bibitem{}  A.M. Dare, L. Chen, A-M. Tremblay \textit{Phys.Rev.}\textbf{\ B49%
} 4106 (1994)

\bibitem{}  E. Abrahams \textit{(private communications)}

\bibitem{}  L. Yin, S. Chakravarty \textit{Int.J.Mod.Phys.} \textbf{B10} 805
(1996)

\bibitem{}  J. Voit \textit{Phys.Rev.} \textbf{B47} 6740 (1993)

\bibitem{}  R. Shankar \textit{Rev.Mod.Phys.}\textbf{\ 66} 129 (1994)

\bibitem{}  S.R. White \textit{Phys.Rev.} \textbf{B44} 4670 (1991)

\bibitem{}  N. Bulut, D.J. Scalapino, S.R. White \textit{Phys.Rev.Lett.} 
\textbf{79} 3752 (1997)

\bibitem{}  R. Ramashvilli, P. Coleman \textit{Phys.Rev.Lett.} \textbf{79}
3752 (1997)

\bibitem{}  A. Rosh, A. Schroder, O. Stockert, H.v. Lohneysen \textit{%
Phys.Rev.Lett.}\textbf{\ 79} 159 (1997)

\bibitem{}  Y.M. Vilk, L. Chen, A-M.S. Tremblay \textit{Phys.Rev.} \textbf{%
B49} 13267 (1994)

\bibitem{}  Y.M. Vilk, A-M.S. Tremblay \textit{EuroPhys.Lett.} \textbf{33}
159 (1996)

\bibitem{}  Y.M. Vilk \textit{Phys.Rev.} \textbf{B55} 3870 (1997)

\bibitem{}  Y.M. Vilk, A-M.S. Tremblay \textit{J.Phys. I (Paris)}\textbf{\ 7 
}1309 (1997)

\bibitem{}  A.V. Chubukov \textit{Phys.Rev.} \textbf{B52} 3840 (1995)

\bibitem{}  A.V. Chubukov, J. Schmalian \textit{Phys.Rev.} \textbf{B57}
11089 (1998)

\bibitem{}  A.J. Millis, H. Monien, D. Pines \textit{Phys.Rev.} \textbf{B42}
176 (1990)

\bibitem{}  V. Barzykin, D. Pines \textit{Phys.Rev.}\ \textbf{B52} 13585
(1995)

\bibitem{}  J. Schmalian \textit{cond-mat/9810041}

\bibitem{}  A.J. Millis \textit{Phys.Rev.} \textbf{B48} 7183 (1993)

\bibitem{}  D. Pines \textit{Z.Phys.} \textbf{B103} 129 (1997)

\bibitem{}  S. Sachdev, A.V. Chubukov, A. Sokol \textit{Phys.Rev.} \textbf{%
B51} 14874 (1995)

\bibitem{}  G. Preosti, Y.M. Vilk, M.R. Norman \textit{cond-mat/9808298}

\bibitem{}  S. Marcelja \textit{Phys.Rev.} \textbf{B1} 2351 (1970)

\bibitem{}  P. Hazenfrantz, F.Niedermager \textit{Phys.Lett.}\ \textbf{B268}
239 (1991)

\bibitem{}  R. Ramazashvilli \textit{cond-mat/9901191}

\bibitem{}  A. Mourachkine \textit{cond-mat/9810161}

\bibitem{}  B. Janko, J. Maly, K. Levin \textit{Phys.Rev.} \textbf{B56}
11407 (1997);Q. Chen, I. Kosztin, B.Janko,K. Levin \textit{cond-mat/980714}

\bibitem{}  M.R. Norman, M. Randeria, H. Ding, J.C. Campuzano \textit{%
Phys.Rev.}\textbf{\ B57} 11093 (1998)

\bibitem{}  N. Kristoffel, T. Ord \textit{Physica C}\textbf{\ 298} 37 (1998)

\bibitem{}  O. Tchernyshyov \textit{Phys.Rev.} \textbf{B56} 3372 (1997)

\bibitem{}  I. Tifrea, I. Grosu, M. Crisan \textit{Physica C} \textbf{298}
51 (1998)

\bibitem{}  D.M. Eagles \textit{Phys.Rev.} \textbf{186} 456 (1969)

\bibitem{}  A.J. Legget \textit{J.Phys. (Paris)} \textbf{C7} 19 (1980)

\bibitem{}  P. Nozieres, S.Schimtt-Rink\textit{\ J.Low Temp.Phys.} \textbf{59%
} 195 (1985)

\bibitem{}  R.Fridberg, T.D. Lee \textit{Phys.Rev.} \textbf{B40} 6745 (1989)

\bibitem{}  S. Schimtt-Rink, C.M. Varma, A.E. Ruckenstein \textit{%
Phys.Rev.Lett.} \textbf{63} 445 (1989)

\bibitem{}  M. Randeria, J.M. Duan, L.Y.Shien \textit{Phys.Rev.} \textbf{B41}
327 (1990)

\bibitem{}  M. Crechster, W. Zwerger \textit{Ann.Phys. (Germany)} \textbf{1}
15 (1992)

\bibitem{}  C.A. R. sa de Melo, M. Randeria, J.R. Engelbrecht \textit{%
Phys.Rev.Lett.} \textbf{71} 3203 (1993)

\bibitem{}  R. Cote, A. Griffin \textit{Phys.Rev.} \textbf{B48} 10404 (1993)

\bibitem{}  L. Belkhir, M. Randeria \textit{Phys.Rev.} \textbf{B49} 6829
(1994)

\bibitem{}  G. Ropke\textit{\ Ann.Phys. (Germany)} \textbf{3} 134 (1994)

\bibitem{}  F. Pistolesi, G.C. Strinati \textit{Phys.Rev.} \textbf{B53}
15168 (1996)

\bibitem{}  S. Stintzing, W. Zwerger \textit{Phys.Rev.} \textbf{B56} 9004
(1997)

\bibitem{}  M. Marina, F. Pistolesi, G.C. Strinati \textit{Euro.Phys.J.}%
\textbf{\ B1} 151 (1998)

\bibitem{}  A.M.J. Schakel \textit{cond-mat/9811393}

\bibitem{}  A.M.J. Schakel \textit{''Boulevard of Broken Symmetries''
cond-mat/9805152}

\bibitem{}  A.M.J. Schakel \textit{''Time dependent Ginsburg-Landau theory
of duality'' (to be published )}

\bibitem{}  A.M.J. Schakel \textit{Ph.D.Thesis Univ.of Amsterdam (1989)}

\bibitem{}  I. Tifrea, M. Crisan \textit{Euro.Phys.J.} \textbf{B4} 175 (1998)

%\newpage \ 
\end{thebibliography}

\begin{thebibliography}{}
\bibitem{}  P. Benard et.al. \textit{Phys.Rev.}\textbf{\ B47} 15217 (1993)

\bibitem{}  L. Chen et.al. \textit{Phys.Rev.}\textbf{\ B52 }1152 (1995)

\bibitem{}  M. Crisan, L. Tataru \textit{J.Supercond.}\textbf{\ 8} 341 (1995)

\bibitem{}  Y.M. Vilk, A-M.S. Tremblay \textit{Phys.Rev.}\textbf{\ B49}
13267 (1994)

\bibitem{}  H.van Loheysen J.Phys.\textit{Cond.Matt.} \textbf{8} 9689 (1996)

\bibitem{}  A.Rosch et.al. \textit{Phys.Rev.Lett.} \textbf{79} 159 (1997)

\bibitem{}  A. Schroder et.al. \textit{Phys.Rev.Lett.} \textbf{80} 5623
(1998)

\bibitem{}  O. Stockert et.al. \textit{Phys.Rev.Lett.}\ \textbf{80 }5627
(1998)

\bibitem{}  P.Coleman\ \textit{cond-mat/9809436}

\bibitem{}  J.A. Hertz \textit{Phys.Rev.} \textbf{B14} 1165 (1976)

\bibitem{}  A.J. Millis \textit{Phys.Rev.} \textbf{B48} 7183 (1993)

\bibitem{}  T. Moriya \textit{Spin Fluctuation in Itinerant Electron
Magnetism (Springer Heidelberg, 1985)}

\bibitem{}  R.Ramazashvilli \textit{cond-mat/9901191}

\bibitem{}  D. Belitz, T.R. Kirkpatrik \textit{Phys.Rev.}\textbf{\ B56} 6513
(1997)

%\newpage\ 
\end{thebibliography}
\end{document}